\documentclass[prd,twocolumn,showpacs,preprintnumbers,amsmath,nofootinbib,amssymb]{revtex4}
\usepackage{graphicx}
\usepackage[sort&compress]{natbib}
\usepackage{subfigure}
\usepackage{amsmath}
\usepackage{amsfonts}
\usepackage{cancel}
\usepackage{lmodern,dsfont}
\usepackage{amssymb}
\usepackage{ulem} 
\usepackage[usenames]{color}
\begin{document}
\arraycolsep1.5pt
\newcommand{\Ima}{\textrm{Im}}
\newcommand{\Rea}{\textrm{Re}}
\newcommand{\mev}{\textrm{ MeV}}
\newcommand{\be}{\begin{equation}}
\newcommand{\ee}{\end{equation}}
\newcommand{\ba}{\begin{eqnarray}}
\newcommand{\ea}{\end{eqnarray}}
\newcommand{\gev}{\textrm{ GeV}}
\newcommand{\nn}{{\nonumber}}
\newcommand{\dtres}{d^{\hspace{0.1mm} 3}\hspace{-0.5mm}}
\newcommand{\rts}{ \sqrt s}
\newcommand{\non}{\nonumber \\[2mm]}


\title{Scalar mesons moving in a finite volume and the role of partial wave mixing}


\author{M. D\"oring$^1$, U.-G. Mei{\ss}ner$^{1,2}$, E. Oset$^3$ and A. Rusetsky$^1$}

\affiliation{
$^1$Helmholtz-Institut f\"ur Strahlen- und Kernphysik (Theorie) and Bethe Center
for Theoretical Physics,  Universit\"at Bonn, Nu\ss allee 14-16, D-53115
Bonn, Germany\\
$^2$Forschungszentrum J\"ulich, J\"ulich Center for Hadron Physics, Institut f\"ur
Kernphysik  (IKP-3) and Institute for Advanced Simulation (IAS-4), D-52425
J\"ulich, Germany\\
$^3$Departamento de F\'{\i}sica Te\'orica and IFIC, Centro Mixto Universidad de Valencia-CSIC,
Institutos de Investigaci\'on de Paterna, Aptdo. 22085, 46071 Valencia,
Spain
}

\begin{abstract}
Phase shifts and resonance parameters can be obtained from finite-volume lattice spectra for
interacting pairs of particles, moving with nonzero total momentum.  We present a simple
derivation of the method that is subsequently applied to obtain the  $\pi \pi$ and $\pi K$
phase shifts in the sectors with total isospin $I=0$ and $I=1/2$, respectively.  Considering
different total momenta, one obtains extra data points for a given volume that allow for a
very  efficient extraction of the resonance parameters in the infinite-volume limit.
Corrections due to the mixing of partial waves are provided.  We expect that our results will
help to optimize the strategies in lattice simulations, which aim at an accurate determination
of the scattering and resonance properties.

\end{abstract}



\maketitle


\section{Introduction}
\label{Intro}

  One of the present issues in QCD lattice calculations is the determination of the excited
hadron spectrum. Many efforts are being devoted to this problem lately
\cite{Nakahara:1999vy,Sasaki:2005ap,Mathur:2006bs,Basak:2007kj,Bulava:2010yg,Morningstar:2010ae,Foley:2010te,Alford:2000mm,Kunihiro:2003yj,Okiharu:2005eg,Suganuma:2005ds,Suganuma:2007uv,McNeile:2006nv,Hart:2006ps,Wada:2007cp,Prelovsek:2004jp,Prelovsek:2010zz,Prelovsek:2010kg,Lin:2008pr,Gattringer:2008vj,Engel:2010my,Mahbub:2010me,Menadue:2011pd,Edwards:2011jj,Prelovsek:2011im,Fu:2011xw,Lang:2012sv}.
In the volume-dependent spectrum,  the ``avoided level crossing" is usually taken as a signal of 
a resonance, but this criterion has been shown insufficient for resonances with a large width
\cite{Bernard:2007cm,Bernard:2008ax,Bernard:2009mw,Bernard:2010fp,Doring:2011vk}. For resonances
with a single decay channel, one often uses L\"uscher's approach to extract phase shifts from the
discrete energy levels in the box \cite{Luscher:1986pf,Luscher:1990ux}. The method has been
recently extended to multi-channel
scattering~\cite{He:2005ey,Liu:2005kr,Lage:2009zv,Bernard:2010fp} and to the case with
three-particle intermediate states~\cite{Polejaeva:2012ut,Kreuzer:2012sr}. Moreover, in
Ref.~\cite{Doring:2011vk},  a method based on coupled-channel Unitary Chiral Perturbation Theory
(UChPT), which operates with the full relativistic two-body propagator, has been proposed.  In
the infinite-volume limit, this method is equivalent to L\"uscher's approach, up to
contributions  (kept in Ref. \cite{Doring:2011vk}), which are exponentially suppressed in this
limit. The new method, combining conceptual and technical simplicity, can provide a guideline for
future lattice calculations. 

This method has been extended in Ref.~\cite{Doring:2011ip} for the use in connection with
dynamical coupled-channel  approaches such as the J\"ulich
model~\cite{Doring:2009bi,Doring:2009yv,Doring:2010ap}.  In Ref.~\cite{MartinezTorres:2011pr}
the framework of Ref.~\cite{Doring:2011vk} has been applied for the interaction of the $DK$ and
$\eta D_s$ system, where the $D_{s^*0}(2317)$ resonance is dynamically generated from the
interaction of these particles. The case of the $\kappa$ in the $K \pi,K\eta$ channels is
addressed along the lines of Ref.~\cite{Doring:2011vk} in Ref.~\cite{Doring:2011nd}, together
with the $\sigma(600)$, $K^*(892)$, and $\rho$ resonances. The formalism has also been extended
to the case of the interaction of unstable particles in Ref.~\cite{Roca:2012rx}, to the study of
the $DN$ interaction \cite{Xie:2012pi}, and the $\pi \pi$ interaction in the $\rho$ channel
\cite{Chen:2012rp}.  Pioneering work for the coupled-channel $\bar KN,\,\pi\Sigma$ system and
the $\Lambda(1405)$  in the finite volume has been carried out in Ref.~\cite{Lage:2009zv};  the
lattice levels for the $\Lambda(1405)$ quantum numbers were evaluated in 
Ref.~\cite{Doring:2011ip} using the J\"ulich model, and strategies to determine the two
$\Lambda(1405)$ states from lattice results (c.f. also Ref.~\cite{Menadue:2011pd}) were
discussed in Ref.~\cite{MartinezTorres:2012yi}.
Methods to extract matrix elements of unstable particles from the finite volume
have been recently developed in Ref.~\cite{Bernard:2012bi}.

  The derivation of Refs.~\cite{Luscher:1986pf,Luscher:1990ux} or Ref.~\cite{Doring:2011vk} is
done for a pair of particles with total zero-momentum. The generalization to a moving frame has
been done in
\cite{Rummukainen:1995vs,Kim:2005gf,Bour:2011ef,Beane:2011sc,Davoudi:2011md,Fu:2011xz,Leskovec:2012gb,Dudek:2012gj,Hansen:2012tf,Briceno:2012yi,Kreuzer:2012sr}.
 
   In this study we present an easy derivation of the approach for moving frames, along the
lines of Ref.~\cite{Doring:2011vk}, using fully relativistic propagators and arbitrary masses,
and we apply the method to study the coupled-channel scattering of $\pi \pi$ and $\pi K$ in the
region of the $f_0(600)$ and  $\kappa(800)$ resonances, respectively.
   
   The strategy followed here is to use the chiral unitary approach to generate synthetic lattice
data, which are later on analyzed to extract phase shifts and resonance properties. We
show that the data produced with moving frames are very useful to get the infinite-volume
properties using two box volumes and different total momenta. The study done here permits to
find optimal strategies, concerning which lattice data to use, in order to obtain phase shifts
and resonance properties with maximum precision. 

 
\section{Formalism}


\subsection{Particles in a moving frame}
\label{sec:movs}
In the chiral unitary approach the scattering matrix in coupled channels is given by the
Bethe-Salpeter equation (BSE) in its factorized form

\be
T=[1-VG]^{-1}V= [V^{-1}-G]^{-1},
\label{bse}
\ee
where $V\equiv V^{(ij)}$ is the matrix for the transition potentials between the channels and 
$G$  is a diagonal matrix with the $i^{\rm th}$ element, $G^{(i)}$, given by the loop function
of two propagators, which for two mesons is defined as 
\be
\label{loop}
G^{(i)}=i\,\int\frac{d^4 q}{(2\pi)^4} \,
\frac{1}{(P-q)^2-m_1^2+i\epsilon}\,\frac{1}{q^2-m_2^2+i\epsilon}
\ ,
\ee
where $m_i$ are the masses of the two mesons and $P$ the total  four-momentum of the 
meson-meson system. The factorized form of the BSE implies an on-shell factorization 
of the potential $V$, see Sec.~\ref{sec:pawa} for a discussion.

$V$ and $T$ in Eq. (\ref{bse}) stand for the potential and scattering matrix in the momentum
space. The normalization is such that, for the case of one  channel, we have
\begin{multline}
\frac{1}{2i}\left(e^{2i\delta}-1\right)=e^{i\delta}\sin\delta=\frac{1}{\cot \delta-i}
\\
=-\frac{p}{8\pi\,E}\,T=-\frac{p}{8\pi\,E}\frac{V}{1-VG}=\frac{1}{2i}\,(S-1)\, ,
\label{st}
\end{multline}
which relates $T$ to the phase shift $\delta$ and the $S$-matrix, where $E$ is the total energy
in  the CM and $p$ the momentum, $p= \lambda^{1/2} ( E^2, m_1^2, m_2^2)/(2E)$. For the
two-channel case, corresponding relations can be found in 
Refs.~\cite{Oller:1998hw,Oller:1997ti}.

The loop function in Eq.~(\ref{loop}) needs to be regularized and this can be accomplished
either with dimensional regularization or with a three-momentum cutoff. The equivalence of both
methods was shown in Refs.~\cite{Oller:2000fj,Oller:1998hw}. In the regularization with a
three-momentum cutoff, one first performs the  $q^0$ integration
analytically~\cite{Oller:1997ti}. As a result, one gets 
\ba
&&G^{(i)}=\hspace{-4mm}\int\limits^{|\vec q\,|<q_{\rm max}}\frac{d^3 \vec q}{(2\pi)^3} \,
I^{(i)}(|\vec q\,|),
\label{prop_cont}
\ea
where 
\ba\label{eq:I-oset}
I^{(i)}(|\vec q\,|)&=&\frac{1}{2\omega_1(\vec q\,)\,\omega_2(\vec q\,)}
\frac{\omega_1(\vec q\,)+\omega_2(\vec q\,)}
{E^2-(\omega_1(\vec q\,)+\omega_2(\vec q\,))^2+i\epsilon},\non
\omega_j^2&=&m_j^2+\vec q^{\,2} \ .
\label{prop_contado}
\ea
We would like to stress that other renormalization schemes such as dimensional regularization
are, in general, preferable over the cut-off renormalization. However,  note that for the
extraction of the infinite-volume limit from lattice levels the cut-off dependence cancels as
discussed in Sec.~\ref{sec:recon_kappa}. The treatment of finite volume using dimensional
regularization is done in Ref.~\cite{MartinezTorres:2011pr}.

To obtain the energy levels in the finite box, instead of integrating over the momenta of the
continuum -- with $q$ being a continuous variable as in Eq.~(\ref{prop_cont}) -- one must sum
over the discrete momenta allowed in a finite box of side length $L$ with periodic boundary
conditions. We then have to replace $G$ by  $\tilde G={\rm diag}\,(\tilde G^{(1)},\tilde
G^{(2)})$, where 
\ba
\tilde G^{(i)}&=&\frac{1}{L^3}\sum_{\vec n}^{|\vec q\,|<q_{\rm max}}
I^{(i)}(|\vec q\,|),
\non 
\vec q&=&\frac{2\pi}{L}\,\vec n,
\quad\vec n\in \mathds{Z}^3 \ .
\label{tildeg}
\ea 
This is the procedure followed in \cite{Doring:2011vk}. Here and in the following, we indicate
quantities in the finite volume with a tilde, e.g. $G\to \tilde G$, or $T\to\tilde T$.

The eigenenergies of the box correspond to energies that produce poles in the $\tilde T$
scattering matrix in the finite volume,
\be
\tilde T=[1-V\tilde G]^{-1}V= [V^{-1}-\tilde G]^{-1},
\label{bse_finite}
\ee
i.e. for energies where $\det(1-V\tilde G)=0$.

In the former discussion the integrals and sums are performed in the rest frame of the two
interacting particles. Yet, since in the infinite volume the $G$ function is Lorentz invariant,
see Eq. (\ref{loop}), it suffices to evaluate it in the two-particle rest frame. In another frame
it will take the same value as required by Lorentz invariance. Let us call $q^*\equiv |\vec
q^{\,*}|$ the absolute value of the relative three-momentum in the rest frame of the two
particles,  called center-of-mass frame (CM) in the following. The absolute value of the relative
three-momentum  in a frame where the two-particle system has total momentum $(P^0, \vec{P})$ is
called $q$. The CM energy of the two-particle system will be $\sqrt s$, such that
\be
s\equiv E^2= (P^0)^2-\vec{P}^2.
\label{se}
\ee 
However, for the system moving in the finite volume, Lorentz invariance is broken and hence we
cannot use $\tilde G$ evaluated in the CM, because the discretization condition in the momenta
$\vec{q_1}$ and  $\vec{q_2}=\vec{P}-\vec{q_1}$ of the particles of Eq.~(\ref{tildeg}) must be
transformed to the moving frame. We must write the boost transformation from $q$ to $q^*$.  By
applying the Lorentz transformation from a moving frame with four-momentum $P$ to a frame where
the two particle system is at rest we find 
\be
\vec{q}^{\,*}_{1,2}=\vec q_{1,2} + \left[\left(\frac{P^0}{\sqrt s}-1\right)
\frac{\vec q_{1,2}\cdot\vec P}{|\vec P|^2}-\frac{q^0_{1,2}}{\sqrt s}\right]\vec P.
\label{boosteq}
\ee
Here and in the following, a star indicates a quantity defined in the two-particle rest frame.
 
Demanding that $\vec{q_1^*}+\vec{q_2^*}=0$ enforces $q_1^0+q_2^0=P^0$. We also have the
transformation of the energies 
\be
q_{1,2}^0=\frac{(q_{1,2}^{*0} \sqrt s + \vec q_{1,2}\cdot \vec P)}{P^0}
\label{q0boost}
\ee
and the condition $q_1^0+q_2^0=P^0$ imposes 
\be
q_1^{*0}+q_2^{*0}=\sqrt s
\ee
which on shell gives 
\be
q^{*0}_{1,2}=\frac{s+m_{1,2}^2-m_{2,1}^2}{2 \sqrt s}.
\label{q0cm}
\ee
This, via Eqs.~(\ref{boosteq}) and (\ref{q0boost}), provides then the boost for the off-shell
momenta in the loop, where $\vec{q}$ is arbitrary but the energy is the on-shell one. Only this
prescription ensures $q^{*0}_1=q^{*0}_2$ for two particles of equal mass in the two-particle
rest frame. Since we need the Jacobian of this transformation, it is useful to rewrite
Eq.~(\ref{boosteq}) in terms of the CM energy of the particles and we find
\be
\vec{q}^{\,*}_{1,2}=\vec q_{1,2} + \left[\left(\frac{\sqrt s}{P^0}-1\right)
\frac{\vec q_{1,2}\cdot\vec P}{|\vec P|^2}-\frac{q_{1,2}^{*0}}{P^0}\right]\vec P \ .
\label{boostmisha}
\ee
This equation is the one used in \cite{mishaunstable}. Furthermore we must substitute  $\int
d^3\vec q^{\,*}/(2\pi)^3$ by  $\int d^3\vec q/(2\pi)^3\,\sqrt{s}/P^0$, where the factor $\sqrt
s/P^0$ is  the Jacobian of the transformation, and then replace the integral by the discrete
sum. In summary, we must perform the substitution
\be
\int \frac{d^3\vec q^{\,*}}{(2\pi)^3}I(|\vec q^{\,*}|) \
\longrightarrow \ \tilde G(P)=\frac{1}{L^3}\,\frac{ \sqrt s}{P^0}
   \sum_{\vec n}I(|\vec q^{\,*}(\vec q\,)|),
\label{gboosteds}
\ee
with 
\be
\vec q=\frac{2\pi}{L}\vec n,\quad \vec n\in \mathds{Z}^3~,
\ee
where for both the sum and the integral the limit is $|\vec{q}^{\,*}|< q_{\rm max}$.

Note that in order to have both $\vec{q_1}$ and  $\vec{q_2}=\vec{P}-\vec{q_1}$ fulfilling the
periodic boundary conditions, the momentum $\vec{P}$ must fulfill them, too, and thus we have 
\be
\vec P=\frac{2\pi}{L}\vec N,\quad \vec N\in \mathds{Z}^3~.
\ee

A clarification on the role of the on-shell reduction used in Eq.~(\ref{bse}) is appropriate. 
In general, one does not know how good this approximation is. However, in the infinite-volume
limit, the on-shell approximation of the potential $V$ derived from a Lagrangian delivers  in
many cases a successful description of the phenomenology, see, e.g., the hadronic model we use
in this work~\cite{Oller:1998hw}.   However, fully covariant off-shell unitarized approaches
have been also  developed~\cite{Nieves:2001wt,Borasoy:2007ku,Bruns:2010sv}.  Off-shell
unitarized approaches  in a three-dimensional reduction are realized,  e.g., in dynamical
coupled-channel models~\cite{Doring:2009bi,Doring:2009yv,Doring:2010ap}.

While there are a conceptual differences between on- and off-shell approaches in the infinite
volume limit, it should be clearly stated that, whatever approach is chosen in the infinite
volume limit,  the levels in the finite volume are determined by the on-shell amplitude up to 
exponentially suppressed effects $\sim e^{-L\,M_\pi}$.


\subsection{One-channel analysis}

   The one-channel problem can be easily solved and is very simple, as shown in
\cite{Doring:2011vk}. The $T$ matrix for the infinite volume can be obtained for the energies
which are eigenvalues of the box by ($E=\sqrt s$) 
\be
T(E)=\left(V^{-1}(E)-G(E)\right)^{-1}= \left(\tilde G(P)-G(E)\right)^{-1} \ . 
\label{extracted_1_channel}
\ee
since $\tilde G(P)=V^{-1}(E)$ is the condition for the $\tilde T$ matrix to have a pole for the
finite box. 

Hence we find, in the one-channel case and assuming that only $S$-wave scattering is present, 
\begin{align}
T(E)^{-1}=\lim_{q_{\rm max}\to \infty}
\Bigg[\frac{1}{L^3}\sum_{\vec{n}}^{|\vec{q}^{\,*}|<q_{\rm max}}
     \frac{E}{P^0}I(|\vec q^{\,*}(\vec q\,)|)\non
-\int\limits^{|\vec q^{\,*}|<q_{\rm max}}\frac{d^3\vec q^{\,*}}{(2\pi)^3} 
    I(|\vec q^{\,*}|)\Bigg] \ . \nonumber
\end{align}

This derivation is very simple and the results can be seen to agree with previous ones
\cite{Rummukainen:1995vs,Kim:2005gf,Davoudi:2011md,Fu:2011xz} when one approximates
$I(q^*)$ by 
\be
\label{eq:I-luescher}
I(q^*) \to \frac{1}{2E}\,\frac{1}{p^2-(\vec {q}^{\,*})^2+i\epsilon}
\ee
where  $p=\lambda^{1/2}(E^2,m_1^2,m_2^2)/(2E)$ and $\lambda(x,y,z)$ stands for  the K\"all\'en
triangle function. It is seen that, with this replacement, our expressions agree with those in
the L\"uscher framework. Furthermore, summing the difference of Eq.~(\ref{eq:I-oset}) and
Eq.~(\ref{eq:I-luescher}) over the momenta, it is immediately seen that the finite-volume
corrections to this quantity are exponentially suppressed. Consequently, the present approach is
equivalent to L\"uscher's approach up to the exponentially suppressed terms for large volumes.
Note however that, for moderately large volumes, these exponentially suppressed terms can be
important numerically, see the discussion in Ref. \cite{Doring:2011vk}.  One should realize,
however, that for values of $LM_\pi$ where these terms play a role,  there are many other
exponentially suppressed corrections.

We would like to stress that the formal dependence on the potential $V$ cancels in 
Eq.~(\ref{extracted_1_channel}) -- in other words, Eq.~(\ref{extracted_1_channel}) contains $T$ and does not contain
$V$ or $G$ individually. 
This was expected from the beginning, because the potential $V$ 
is not an observable and
depends on the cut-off chosen for $G$. For the case of moving frames and partial wave mixing,
discussed in the following, we will find exactly the same behavior:
the measured lattice levels depend only on the $T$-matrix in the infinite volume and and functions $\tilde G-G$, which are
cutoff-independent.


\subsection{Partial wave decomposition in a finite volume}
\label{sec:pawa}

To determine the mixing of partial waves, consider first the case without boost, i.e. $\vec
P=\vec 0$. At the end of this section, the formalism is generalized to moving frames and
multiple channels.

We use the spherical harmonics $Y_{\ell m}$ with the normalization
\be
\int_0^{2\pi}d\phi\int_0^\pi \sin\theta\,d\theta 
\,Y_{\ell m}(\theta,\phi)Y^*_{\ell' m'}(\theta,\phi)=\delta_{\ell\ell'}\delta_{mm'}\, ,
\label{norm}
\ee
and further define
\ba\label{curlyY}
{\cal Y}_{\ell m}(\vec p\,)&=&p^\ell Y_{\ell m}(\theta,\phi)=p^\ell Y_{\ell m}(\hat p)\, ,
\nonumber\\
\vec p&=&p\,(\sin\theta\cos\phi,\sin\theta\sin\phi,\cos\theta)=p\,\hat p\, .
\ea
The potential $V$ in the Bethe-Salpeter equation (\ref{bse}) is the same in a finite and in the
infinite volume, $V=\tilde V$. Its partial-wave expansion takes the standard form
\ba
V(\vec p, \vec p\,')&=&4\pi\sum_{\ell,m}\,{\cal Y}_{\ell m}(\vec p\,)\,v_\ell(p,p')
\,{\cal Y}^*_{\ell m}(\vec p\,') \ .
\label{curlyV}
\ea
However, the partial-wave expansion of the scattering amplitude $\tilde T$ is different in a
finite volume, because here the  rotational symmetry is broken down to cubic symmetry. As a
result,
\ba
\tilde T(\vec p, \vec p\,')&=&4\pi\sum_{\substack{\ell,m\\ \ell' m'}}
\,{\cal Y}_{\ell m}(\vec p\,)\, t_{\ell m,\ell'm'}(p,p')\,{\cal Y}^*_{\ell' m'}(\vec p\,')\, .
\label{vt_pw}
\ea
In the above expressions, $v_\ell$ and $ t_{\ell m,\ell'm'}$ depend only on $p^2=\vec p^{\,2}$
and  $(p')^2= (\vec p\,')^2$. Note that the threshold behavior of the amplitude is hidden in 
the function ${\cal Y}_{\ell m}(\vec p\,)\sim p^{\ell}$. 

In the infinite-volume limit, the rotational symmetry is restored and the Wigner-Eckart theorem
guarantees that $ t$ is diagonal both in  $\ell$ and $m$, 
\ba
 t_{\ell m,\ell'm'}= t_{\ell}\delta_{\ell\ell'}\delta_{mm'} \ .
\ea
In contrast, in the finite volume one obtains
\begin{multline}
 t_{\ell m,\ell'm'}(p,p')=v_\ell(p,p')\,\delta_{\ell\ell'}\delta_{mm'}
+\frac{4\pi}{L^3}\,\sum_{\vec n}\sum_{ \ell'' m''}\\
\times
v_\ell(p,q)\,{\cal Y}^*_{\ell m}(\vec q\,)\,I(q)\,
{\cal Y}_{\ell'' m''}(\vec q\,)\, t_{\ell''m'',\ell'm'}(q,p') 
\label{full}
\end{multline}
with $I(q)$ from Eq.~(\ref{prop_contado}). This relation is obtained straightforwardly by inserting $V$
from  Eq.~(\ref{curlyV}) and $\tilde T$ from Eq.~(\ref{vt_pw}) in Eq.~(\ref{bse_finite}).

Next, note that, since $v_\ell(p,p')$ and  $ t_{\ell m,\ell'm'}(p,p')$ depend only on
$p^2,(p')^2$, the use of the regular summation theorem~\cite{Luscher:1986pf} is justified. 
Further, 
the
argument $q$ in these functions [see Eq.~(\ref{full})]
can be replaced by the on-shell value $q^{\rm on}$, which is determined from the zero of the
energy denominator in Eq.~(\ref{prop_contado}). Finally, using Eq.~(\ref{curlyY}), one gets
\begin{multline}
\label{linear-finvol}
\tilde T_{\ell m,\ell'm'}(p,p')=V_\ell(p,p')\,\delta_{\ell\ell'}\delta_{mm'}\\
+\sum_{ \ell'' m''}V_\ell(p,q^{\rm on})\tilde G_{\ell m,\ell''m''}(q^{\rm on})
\tilde T_{\ell''m'',\ell'm'}(q^{\rm on},p')\, ,
\end{multline}
where
\ba
\tilde T_{\ell m,\ell'm'}(p,p')&=&p^\ell t_{\ell m,\ell'm'}(p,p')(p')^{\ell'}\, ,\nonumber\\
V_\ell(p,p')&=&p^\ell v_\ell(p,p')(p')^\ell \ ,
\ea
and 
\ba
\label{G-finite}
\tilde G_{\ell m,\ell'm'}(q^{\rm on})=\frac{4\pi}{L^3}\sum_{\vec n}\,
\biggl(\frac{q}{q^{\rm on}}\biggr)^{\ell+\ell'}\,
Y^*_{\ell m}(\hat q)\,I(q)\, Y_{\ell' m'}(\hat q)\, .\nonumber\\
\ea
The factor $(q/q^{\rm on})^{\ell+\ell'}$ can be
replaced by 1, when $\ell+\ell'$ is even, and by $q/q^{\rm on}$
otherwise~\cite{Polejaeva:2012ut}, i.e. 
\ba
\label{replace}
(q/q^{\rm on})^{\ell+\ell'}\to (q/q^{\rm on})^k\, ,\nonumber\\
k=0,1\quad\mbox{for}\quad \ell+\ell'=\mbox{even, odd}\, .
\ea
Note that only even $\ell+\ell'$ lead to non-zero contributions for $\vec P=\vec 0$,
as well as in case of equal-mass particle scattering. 

Note also that if the factor $(q/q^{\rm on})$ is neglected one obtains expressions that are
different from the original L\"uscher approach~\cite{Luscher:1986pf,Luscher:1990ux} not by
exponentially suppressed terms, but terms suppressed as $1/L^4$~\cite{Polejaeva:2012ut}. In any
case, we have checked numerically that effects for the results of this study, coming from the
$(q/q^{\rm on})$ factor,  are very small for the considered realistic box sizes. 

An important remark is in order. At first glance, it seems that there is an ambiguity in the
choice of $k$ in Eq.~(\ref{replace}). Note, however, that this problem arose because one insisted
on the {\it on-shell} prescription in the infinite volume limit. The above choice ensures that
our finite-volume expressions are compatible with L\"uscher's
approach~\cite{Luscher:1986pf,Luscher:1990ux} up to exponentially suppressed terms, and in the
infinite-volume limit they are also compatible with the on-shell prescription. In particular, the
quantity $\tilde G$ in the infinite-volume limit is replaced by
\ba\label{G-infinite}
\tilde G_{\ell m,\ell'm'}(q^{\rm on})&\xrightarrow[L\to\infty]{}&
\delta_{\ell\ell'}\delta_{mm'}\int\frac{d^3\vec q}{(2\pi)^3}\,I(q)\non
&=&\delta_{\ell\ell'}\delta_{mm'}\,G \ ,
\ea
with $G$ from Eq.~(\ref{prop_cont}) (channel index omitted here). The Bethe-Salpeter equation
in the infinite-volume limit takes the simple form given in Eq.~(\ref{bse}), i.e.
\be
T_{\ell}(p,p')=V_\ell(p,p')+V_\ell(p,q^{\rm on})\, G(q^{\rm on})\, 
T_{\ell}(q^{\rm on},p')
\label{pwa1}
\ee
where the arguments are quoted explicitly. Finally,  recall that the expressions in
Eqs.~(\ref{G-finite}) and  (\ref{G-infinite}) are defined with an implicit momentum cutoff at
$q=q_{\rm max}$.


\subsection{Partial wave mixing with boost and multiple channels}
\label{sec:combi}
In Sec.~\ref{sec:pawa}, we have considered the partial-wave expansion for a zero total momentum,
$\vec P=\vec 0$, of the two particles. Nothing changes conceptually if we consider moving frames
instead, and the formalism of Sec.~\ref{sec:movs} can be applied. In this case, the quantity
$\tilde G$ is given by
\ba
\tilde G_{\ell m,\ell''m''}=\frac{4\pi}{L^3}\sum_{\vec n}^{|\vec q^{\,*}|<q_{\rm max}}
\frac{E}{P_0}
\biggl(\frac{q^*}{q^{\rm on*}}\biggr)^k\nonumber\\
\times
Y^*_{\ell m}(\hat q^*)\,Y_{\ell''m''}(\hat q^*)\,I(q^*) \ .
\label{glm}
\ea
Here, $\vec q^{\,*}$ is given by Eq.~(\ref{boostmisha}),  $E\equiv\sqrt{s}$ is from
Eq.~(\ref{se}), $k$ from Eq.~(\ref{replace}), and  $\tilde G_{\ell m,\ell''m''}$ is obviously the
generalization of $\tilde G$ from Eq.~(\ref{gboosteds}) to higher partial waves. Note that, in
case of particles with different masses, the states with even values of $\ell$  and odd values of
$\ell'$ can mix in a moving frame. In this case, as discussed above,  we choose $k=1$.

In the large-$L$ limit, up to the terms exponentially suppressed
in $L$, the quantity $\tilde G_{\ell m,\ell'' m''}$ can be expressed in terms of L\"uscher's zeta-functions. Namely,
\ba\label{eq:mapping}
\tilde G_{\ell m,\ell'' m''}-\delta_{\ell \ell''}\delta_{mm''}G
=-\frac{q^{{\sf on}*}}{8\pi\sqrt{s}}\,i^{\ell-\ell''}\,{\cal M}_{\ell m,\ell'' m''}\, ,
\ea
where ${\cal M}_{\ell m,\ell'' m''}$ is given, e.g., by Eq.~(39) of 
Ref.~\cite{Gockeler:2012yj}. It is a linear combination of the  L\"uscher zeta-functions in the moving frame.

The discrete levels in a finite volume emerge at the energies where the  determinant of the
linear equation~(\ref{linear-finvol}) vanishes,
\ba
\label{det0}
\det\big(\delta_{\ell\ell'}\delta_{mm'}-
V_\ell(q^{\rm on},q^{\rm on})\tilde G_{\ell m,\ell'm'}(q^{\rm on})\big)=0\, .
\ea
Here, $\tilde G$ is a matrix with a row (column) index given by all combinations of $\ell,
m$ ($\ell', m'$).

Finally, all above formulae refer to the single-channel case. In case of  multiple coupled
channels, both the potential $V$ and the quantity $\tilde G$ should be considered as matrices
in channel space.  In that case, $V$, which is diagonal in the space of partial waves, obtains
additional channel indices $V_\ell\to V^{(ij)}_\ell$. The quantity $\tilde G$, that has
non-diagonal elements in the space of partial waves due to the mixing, becomes a diagonal
matrix in channel space, $\tilde G_{\ell m,\ell'm'}\to\tilde G_{\ell m,\ell'm'}^{(i)}$, c.f.
also Eq.~(\ref{loop}). 

In the following, along with UChPT, we shall use the Inverse Amplitude Method (IAM) in the
version of Ref.~\cite{Oller:1998hw} in the finite volume, as has been done already in
Ref.~\cite{Doring:2011nd} for the case $\vec P=0$. 
The IAM method exploits unitarity, which in one channel states that  Im~$T^{-1}= p/(8 \pi
E)$ (see Eq.~(\ref{st})). Denoting by $V^{[2]}$ and $V^{[4]}$ the second- and fourth-order chiral
potentials, and noting that the amplitude at lowest order, $V^{[2]}\equiv T^{[2]}$, has no
imaginary part, a  dispersion relation is made for the function $(V^{[2]})^2/T$, where the 
imaginary part is thus known analytically.  It leads to a simple relationship $T^{\rm IAM}=
(T^{[2]})^2/(T^{[2]}-T^{[4]})$, where $T^{[2]}$, and $T^{[4]}$  are the amplitudes at lowest
order and next-to-lowest order, respectively~\cite{Dobado:1996ps}. We can then recast these
results in terms of Eq.~(\ref{bse}), redefining a potential $V^{[4]}= T^{[4]}- V^{[2]}\, G\,
V^{[2]}$. The equivalent potential to be used in Eq.~(\ref{bse}) or (\ref{det0}) to obtain the
results of the IAM amplitude is then given by
\ba
\label{replaceia2}
V^{\rm IAM}_\ell=\left(1-V^{[4]}(V^{[2]})^{-1}\right)^{-1}V^{[2]} \ .
\ea
Here,
$V^{[2]}\equiv(V^{[2]})^{(ij)}$ and $V^{[4]}\equiv(V^{[4]})^{(ij)}$ are matrices in channel
space. For $V^{[4]}$, only the polynomial terms are considered like in Ref.~\cite{Oller:1998hw}.
The method derived here could be also extended to the loop calculations of
Refs.~\cite{Nieves:2001de,GomezNicola:2001as,Hanhart:2008mx,Nebreda:2010wv}.

For illustration, an explicit example for the emerging structure is quoted. With
$V_\ell^{(ij)}\equiv (V^{\rm IAM}_\ell)^{(ij)}$, we consider the mixing of $S$- and $P$-waves
in the coupled channels $\pi K$ and $\eta K$. Then, $V$ and $\tilde G$ in Eq.~(\ref{det0}) are
given by
\ba
\label{vvvv}
&&V^{(ij)}_\ell=\non
&&\begin{pmatrix}
V_{0}^{(11)}&V_{0}^{(12)}&0&0&0&0&0&0\\
V_{0}^{(21)}&V_{0}^{(22)}&0&0&0&0&0&0\\
0&0&V_{1}^{(11)}&V_{1}^{(12)}&0&0&0&0\\
0&0&V_{1}^{(21)}&V_{1}^{(22)}&0&0&0&0\\
0&0&0&0&V_{1}^{(11)}&V_{1}^{(12)}&0&0\\
0&0&0&0&V_{1}^{(21)}&V_{1}^{(22)}&0&0\\
0&0&0&0&0&0&V_{1}^{(11)}&V_{1}^{(12)}\\
0&0&0&0&0&0&V_{1}^{(21)}&V_{1}^{(22)}\non
\end{pmatrix}\non
\ea
and
\begin{widetext}
\ba
\tilde G^{(i)}_{\ell m,\ell' m'}=
\begin{pmatrix}
\tilde G_{00,00}^{(1)}&0&\tilde G_{00,1-1}^{(1)}&0&\tilde G_{00,10}^{(1)}&0&\tilde G_{00,11}^{(1)}&0\\
0&\tilde G_{00,00}^{(2)}&0&\tilde G_{00,1-1}^{(2)}&0&\tilde G_{00,10}^{(2)}&0&\tilde G_{00,11}^{(2)}\\
\tilde G_{1-1,00}^{(1)}&0&\tilde G_{1-1,1-1}^{(1)}&0&\tilde G_{1-1,10}^{(1)}&0&\tilde G_{1-1,11}^{(1)}&0\\
0&\tilde G_{1-1,00}^{(2)}&0&\tilde G_{1-1,1-1}^{(2)}&0&\tilde G_{1-1,10}^{(2)}&0&\tilde G_{1-1,11}^{(2)}\\
\tilde G_{10,00}^{(1)}&0&\tilde G_{10,1-1}^{(1)}&0&\tilde G_{10,10}^{(1)}&0&\tilde G_{10,11}^{(1)}&0\\
0&\tilde G_{10,00}^{(2)}&0&\tilde G_{10,1-1}^{(2)}&0&\tilde G_{10,10}^{(2)}&0&\tilde G_{10,11}^{(2)}\\
\tilde G_{11,00}^{(1)}&0&\tilde G_{11,1-1}^{(1)}&0&\tilde G_{11,10}^{(1)}&0&\tilde G_{11,11}^{(1)}&0\\
0&\tilde G_{11,00}^{(2)}&0&\tilde G_{11,1-1}^{(2)}&0&\tilde G_{11,10}^{(2)}&0&\tilde G_{11,11}^{(2)}\\
\end{pmatrix}.\non
\label{gmat}
\ea
\end{widetext}


\subsection{Symmetries of Eq.~(\ref{det0})}
\label{akakisolve}

As mentioned above, rotational symmetry is broken down to the cubic group on a finite lattice,
and a mixing of different partial waves occurs. In moving frames, the symmetry is further broken
down to the different subgroups  of the cubic group (different little groups). Using  symmetry
arguments, it is possible to carry out a partial diagonalization of Eq.~(\ref{det0}), as well as
to construct the operators that project out the spectra corresponding to the different
irreducible representations of the little groups. A full-fledged analysis of the problem has been
carried out recently~\cite{Gockeler:2012yj}. For earlier work on the subject, see, e.g.,
Refs.~\cite{Luscher:1990ux,Rummukainen:1995vs,Bernard:2008ax,Luu:2011ep,Fu:2011xz,Leskovec:2012gb} 
and references therein. Below, we use the results of Ref.~\cite{Gockeler:2012yj}, in case of the
meson-meson  scattering in $S$, $P$, $D$ waves, in order to attribute the emerging structures to
the different little groups.


\subsubsection{The case ${\vec P}=(2\pi/L)\,(0,0,0)$}

The partial waves $S,P,D$ do not mix -- the pertinent part of the  matrix $\tilde G$ is diagonal
in $\ell,\ell'$ and all entries with $m-m'\neq 0 \mod 4$  vanish. Non-vanishing elements are:
$\tilde G_{00,00}\, ,$
$\tilde G_{11,11}=\tilde G_{10,10}=\tilde G_{1-1,1-1}\, ,$ 
$\tilde G_{21,21}=\tilde G_{2-1,2-1}\, ,$ 
$\tilde G_{20,20}\, ,$
$\tilde G_{22,22}=\tilde G_{2-2,2-2}\, ,$ 
$\tilde G_{22,2-2}=\tilde G_{2-2,22}=(\tilde G_{20,20}-\tilde G_{21,21})/2\, .$

Below we list the irreducible representations for a given $\ell$ and the pertinent equations for
the determination of the energy levels~\cite{Luscher:1990ux}, obtained from Eq.~(\ref{det0}): 
\begin{align}
\ell=0,\quad A_1^+:&&
1-V_0\tilde G_{00,00}=0\, ,\non
\ell=1,\quad T_1^-:&&
1-V_1\tilde G_{10,10}=0\, ,\non
\ell=2,\quad E^+:&&
1-V_2\tilde G_{21,21}=0\, ,\non
\ell=2,\quad T_2^+:&&
1-V_2\tilde G_{20,20}=0\, .
\label{p000}
\end{align}
Note that for more than one channel, $V_\ell$ and $\tilde G_{\ell m,\ell'm'}$ in these
equations can be considered as matrices in channel space. See also Sec.~\ref{sec:combi}. The
energy levels are then given by the zeros of the corresponding determinants. For the cases of
boosts with $\vec P\neq 0$, one has to be slightly more  careful as the order of the matrices
matter, and the formulae of the following sections are derived for the one-channel case. An
example of partial wave mixing with multiple channels is explicitly evaluated in
Sec.~\ref{sec:spectrumsigma}.


\subsubsection{The case ${\vec P}=(2\pi/L)\,(0,0,1)$}

In the unequal mass case, we restrict ourselves to $S,P$ waves only. This corresponds to the
partial wave mixing studied for the $\kappa(800),\,K^*(892)$ system in Sec.~\ref{sec:kappa}. The
non-zero matrix elements of $\tilde G$ are:
$\tilde G_{00,00}\, ,$
$\tilde G_{11,11}=\tilde G_{1-1,1-1}\, ,$
$\tilde G_{10,10}\, ,$
$\tilde G_{00,10}=\tilde G_{10,00}\, .$

The irreducible representations and the equations for the determination of the energy levels
are given by solving Eq.~(\ref{det0}):
\begin{align}
\ell&=0,1, & A_1:\hspace{0.7cm}&
(1-V_0\tilde G_{00,00})(1-V_1\tilde G_{10,10})&&\non
&&&-V_0V_1\tilde G_{00,10}^2=0\, ,&&\non
\ell&=1, & E:\hspace{0.7cm}&
1-V_1\tilde G_{11,11}=0\, .&&
\label{p001sp}
\end{align}
Note that here and in the following, the $\tilde G_{\ell m,\ell'm'}$ can be mapped to
linear combinations of the L\"uscher zeta-functions or the quantities 
$w_{\ell'' m''}$ used, e.g., in
Refs.~\cite{Leskovec:2012gb,Fu:2011xz} (see Eq.~(\ref{eq:mapping})).

We shall need $D$-waves in the equal-mass case only. In particular, we will study the mixing of
$D$-waves with $S$-waves for the $\sigma(600)$ isoscalar in Sec.~\ref{sec:sigma}. For equal
masses, $\tilde G_{00,10}=0$ and the mixing between the $S$- and $P$-waves vanishes. The mixing
vanishes anyway for $\pi\pi$ as $L+I$ =even, and the $P$-wave has isospin 1. 

Additional non-zero matrix elements are: 
$\tilde G_{00,20}=\tilde G_{20,00}\, ,$
$\tilde G_{20,20}\, ,$
$\tilde G_{22,22}=\tilde G_{2-2,2-2}\, ,$
$\tilde G_{21,21}=\tilde G_{2-1,2-1}\, ,$
$\tilde G_{22,2-2}=\tilde G_{2-2,22}\, .$

The irreducible representations and the equations for the determination of the energy levels
are:
\begin{align}
\ell&=0,2, & A_1^+:\hspace{0.7cm}&
(1-V_0\tilde G_{00,00})(1-V_2\tilde G_{20,20})\non
&&&-V_0V_2\tilde G_{00,20}^2=0\, ,\non
\ell&=2, & E^+:\hspace{0.7cm}&
1-V_2\tilde G_{21,21}=0\, ,\non
\ell&=2, & B_1^+:\hspace{0.7cm}&
1-V_2(\tilde G_{22,22}+\tilde G_{22,2-2})=0\, ,\non
\ell&=2, & B_2^+:\hspace{0.7cm}&
1-V_2(\tilde G_{22,22}-\tilde G_{22,2-2})=0\, .
\label{p001sd}
\end{align}


\subsubsection{The case ${\vec P}=(2\pi/L)\,(1,1,0)$}

For the unequal mass case, we again consider $S,P$ waves only. The non-zero matrix elements of
$\tilde G$ are:
$\tilde G_{00,00}\, ,$
$\tilde G_{00,1-1}\, ,$
$\tilde G_{00,11}=-i \tilde G_{00,1-1}\, ,$
$\tilde G_{1-1,00}=i \tilde G_{00,1-1}\, ,$
$\tilde G_{11,00}=-\tilde G_{00,1-1}\, ,$
$\tilde G_{10,10}\, ,$
$\tilde G_{11,11}=\tilde G_{1-1,1-1}\, ,$
$\tilde G_{11,1-1}=-\tilde G_{1-1,11}\, .$
The irreducible representations and the equations for the determination of the energy levels
are:
\begin{align}
\ell&=0,1,& A_1:\hspace{0.2cm}&
(1-V_0\tilde G_{00,00})\non
&&&\times(1-V_1(\tilde G_{1-1,1-1}-i\tilde G_{11,1-1}))\non
&&& -V_0V_1(\tilde G_{00,11}-\tilde G_{00,1-1})^2=0\, ,\non
\ell&=1,& B_1:\hspace{0.2cm}&
1-V_1\tilde G_{10,10}=0\, ,\non
\ell&=1,& B_2:\hspace{0.2cm}&
1-V_1(\tilde G_{1-1,1-1}+i\tilde G_{11,1-1})=0\, .
\end{align}
The mixing of $S$- and $D$-waves is again considered in the equal-mass case. The
irreducible representations and the equations for the determination of the energy levels are:
\begin{align}
\ell&=0,2,& A_1^+:\hspace{0.2cm}&
(1-V_0\tilde G_{00,00})\Big[(1-V_2\tilde G_{20,20})
\non &&&
\times(1-V_2(\tilde G_{2-2,2-2}-\tilde G_{2-2,22}))
\non &&&
 +2V_2^2\tilde G_{2-2,20}^2\Big]
 \non &&&
+2(1-V_2\tilde G_{20,20})V_0V_2\tilde G_{00,2-2}^2
\non &&&
-\tilde G_{00,20}V_0V_2\Big[\tilde G_{00,20}
\non &&&
\times(1-V_2\tilde G_{2-2,2-2}
 +V_2\tilde G_{2-2,22})
 \non &&&
 +4V_2\tilde G_{00,2-2}\tilde G_{2-2,20} \Big]=0\, ,\non
\ell&=2,& A_2^+:\hspace{0.2cm}&
1-V_2(\tilde G_{2-1,2-1}-i\tilde G_{2-1,21})=0\, ,\non
\ell&=2,& B_1^+:\hspace{0.2cm}&
1-V_2(\tilde G_{2-1,2-1}+i\tilde G_{2-1,21})=0\, ,\non
\ell&=2,& B_2^+:\hspace{0.2cm}&
1-V_2(\tilde G_{2-2,2-2}+\tilde G_{2-2,22})=0\, .
\end{align}


\subsubsection{The case ${\vec P}=(2\pi/L)\,(1,1,1)$}

For the unequal mass case, we consider $S,P$ waves only. The non-zero matrix elements of
$\tilde G$ are:
$\tilde G_{00,00}\, ,$
$\tilde G_{00,1-1}=-\tilde G_{11,00}=2^{-1/2}(1-i)\tilde G_{00,10}\, ,$
$\tilde G_{1-1,00}=-\tilde G_{00,11}=2^{-1/2}(1+i)\tilde G_{00,10}\, .$
$\tilde G_{00,10}=\tilde G_{10,00}\, ,$
$\tilde G_{1-1,1-1}=\tilde G_{10,10}=\tilde G_{11,11}\, ,$
$\tilde G_{11,1-1}=-\tilde G_{1-1,11}\, ,$
$\tilde G_{1-1,10}=-\tilde G_{10,11}=2^{-1/2}(1-i)\tilde G_{11,1-1}\, ,$
$\tilde G_{11,10}=-\tilde G_{10,1-1}=2^{-1/2}(1+i)\tilde G_{11,1-1}\, .$
The irreducible representations and the equations for the determination of the energy levels
are:
\begin{align}
\ell&=0,1,& A_1:\hspace{0.3cm}&
(1-V_1(\tilde G_{1-1,1-1}+2i\tilde G_{1-1,11}))
\non &&&
\times(1-V_0\tilde G_{00,00})
-3V_0V_1\tilde G_{00,10}^2=0\, , \non
\ell&=1,& E:\hspace{0.3cm}&
1-V_1(\tilde G_{1-1,1-1}-i\tilde G_{1-1,11})=0\, .
\end{align}

The mixing of $S$- and $D$-waves is considered in the equal-mass case. The irreducible
representations and the equations for the determination of the energy levels are:
\begin{align}
\ell&=0,2,& A_1^+:\hspace{0.4cm}&
\frac{V_0 V_2}{10}\,\left(4\sqrt{3}\tilde G_{2-2,20}+3\sqrt{2}\tilde G_{2-1,21}\right)^2
\non &&&
+ (1-V_2(\tilde G_{2-1,2-1}+2i\tilde G_{2-1,21}))
\non &&&
\times(1-V_0\tilde G_{00,00})=0
\, ,\non
\ell&=2,& E^+:\hspace{0.4cm}&
(1-V_2\tilde G_{20,20})
\non &&&
(1-V_2(\tilde G_{2-1,2-1}-i\tilde G_{2-1,21}))
\non &&&
-6V_2^2\,\tilde G_{2-1,20} \tilde G_{20,2-1}=0\, .
\end{align}
As mentioned before, our results completely agree with those of Ref.~\cite{Gockeler:2012yj} in all partial waves. Further,
note that in Ref.~\cite{Leskovec:2012gb}, the equations are derived for the cases $\vec
P=(2\pi/L) (0,0,1)$ and $\vec P=(2\pi/L) (1,1,0)$, only $S$ and $P$ waves retained. We have
checked that our results exactly agree with the results of Ref.~\cite{Leskovec:2012gb} for these
cases (the representations $B_1,B_2$ for $\vec P=(2\pi/L) (1,1,0)$
are denoted by $B_3,B_2$ in Ref.~\cite{Leskovec:2012gb}). 
In Ref.~\cite{Fu:2011xz}, results for the boosts $P=(2\pi/L) (0,0,1)$ and $\vec P=(2\pi/L)
(1,1,0)$ have been obtained, and in Ref.~\cite{Dudek:2012gj} results equivalent to the present
ones for the $SD$-wave case were determined.


\subsection{Calculation of the energy levels}
\label{michasolve}

In practical calculations, we found it convenient to consider Eq.~(\ref{det0}) in a rotated
frame. We namely rotate the $\hat z$ direction of the coordinate system, in which
$(\theta,\phi)$ of $\vec q^{\,*}$ are measured, into the direction of the boost vector $\vec P$.
To rotate $\hat z$ actively into the $\vec P$ direction -- the latter given by
$(\theta_P,\,\phi_P)$ -- one has different choices; we choose here first a rotation around the
$\hat x$-axis by $-\theta_P$, so that the new $\hat z\,'$ vector is in the $yz$ plane, and then
a rotation around the original $\hat z$-axis by $\phi_P-\pi/2$ so that $\hat z\,'' ||\vec P$.
The coordinates $\vec q^{\,*}\,''$ of a vector $\vec q^{\,*}$, in the rotated coordinate system
with $z$-axis $\hat z\,''$, are then given by the inverse of these (non-commuting) rotations.
With standard rotation matrices $R_i,\,i=x,y,z$, the new Cartesian coordinates read
\ba
\vec q^{\,*}\,''&=&R\vec q^{\,*}
  =\left(R_z\left(\phi_P-\frac{\pi}{2}\right)
     R_x\left(-\theta_P\right)\right)^{-1}\vec q^{\,*} \ ,\non
R&=&\begin{pmatrix}
\sin\phi_P		& -\cos\phi_P 			& 0		\\
\cos\theta_P\cos\phi_P	& \cos\theta_P\sin\phi_P	& -\sin\theta_P	\\
\sin\theta_P\cos\phi_P	& \sin\theta_P\sin\phi_P	& \cos\theta_P	
\end{pmatrix} \ .
\ea
where
\begin{align}
\theta_P&=0,			& \phi_P&=0 	& \text{ for } P=(2\pi/L) (0,0,1) \ ,\non
\theta_P&=\pi/2,		& \phi_P&=\pi/4	& \text{ for } P=(2\pi/L) (1,1,0) \ ,\non
\theta_P&=\arctan\sqrt{2}, 	& \phi_P&=\pi/4 & \text{ for } P=(2\pi/L) (1,1,1) \ .\non
\end{align}
We would like to stress that all calculations in this paper have been 
carried out in the rotated frame. The formulae listed in the 
previous section merely serve to demonstrate the equivalence of the 
present approach to that of Ref.~\cite{Gockeler:2012yj}.

From now on we will only consider the $\tilde G$ functions in the rotated coordinate system
(compare to Eq.~(\ref{glm})),
\ba
\tilde G^R_{\ell m,\ell'm'}=\frac{4\pi}{L^3}\sum_{\vec n}^{|\vec q^{\,*}|<q_{\rm max}}
\frac{E}{P_0}
\biggl(\frac{q^*}{q^{\rm on*}}\biggr)^k\nonumber\\
\times
Y^*_{\ell m}(\hat q^*\,''(\hat q^*))\,Y_{\ell'm'}(\hat q^*\,''(\hat q^*))\,I(q^*) \ .
\label{glmrot}
\ea
In particular, with the rotation of the coordinate system the zeros of the determinant, i.e.
the levels, remain unchanged and the matrix of Eq.~(\ref{det0}) becomes block-diagonal.

For the boosts $\vec P=(2\pi/L)(0,0,0)$, $(2\pi/L) (0,0,1)$, $(2\pi/L)(0,1,0)$,
$(2\pi/L) (1,0,0)$, $(2\pi/L) (0,0,2)$, $\cdots$,  Eqs.~(\ref{p000}), (\ref{p001sp}),
(\ref{p001sd}) remain of course valid with $\tilde G^R$ from Eq.~(\ref{glmrot}), because for
these boosts $\tilde G^R_{\ell m,\ell'm'}=\tilde G_{\ell m,\ell'm'}$. For the boost
$\vec{P}=(2\pi/L) (1,1,0)$, we obtain:
\begin{align}
\ell&=0,1,& A_1:\hspace{0.4cm} &
(1-V_0\tilde G^R_{00,00})(1-V_1\tilde G^R_{10,10})
\non &&&
-V_0V_1(\tilde G^R_{00,10})^2
 =0\, ,\non
\ell&=1,& B_{1}:\hspace{0.4cm} &
1-V_1(\tilde G^R_{11,11}+\tilde G^R_{11,1-1})=0\, , \non
\ell&=1,& B_{2}:\hspace{0.4cm} &
1-V_1(\tilde G^R_{11,11}-\tilde G^R_{11,1-1})=0\, ,\non
\ell&=2,& A_2^+:\hspace{0.4cm} &
1-V_2(\tilde G^R_{22,2-2}-\tilde G^R_{22,22})=0\, ,\non
\ell&=2,& B_{1}^+:\hspace{0.4cm} &
1-V_2(\tilde G^R_{21,21}+\tilde G^R_{21,2-1})=0\, ,\non
\ell&=2,& B_{2}^+:\hspace{0.4cm} &
1-V_2(\tilde G^R_{21,21}-\tilde G^R_{21,2-1})=0\, ,\non
\ell&=0,2,& A_1^+:\hspace{0.4cm} &
1-V_0 \tilde G^R_{00,00}-a\,V_2 +b\,V_2^2 =0
\label{rot1}
\end{align}
where
\ba
a&=&\tilde G^R_{2-2,2-2}+\tilde G^R_{2-2,22}+\tilde G^R_{20,20}\non
 &+&V_0 \big[2 (\tilde G^R_{00,2-2})^2+(\tilde G^R_{00,20})^2\non
 && -\tilde G^R_{00,00}
   \left(\tilde G^R_{2-2,2-2}+\tilde G^R_{2-2,22}+\tilde G^R_{20,20}\right)\big] \ ,\non
b&=&2 (\tilde G^R_{2-2,20})^2 \left(V_0 \tilde G^R_{00,00}-1\right)\non
    &-&4 V_0 \tilde G^R_{00,2-2} \tilde G^R_{00,20} \tilde G^R_{2-2,20}\non
    &+&\left(\tilde G^R_{2-2,2-2}+\tilde G^R_{2-2,22}\right) \tilde G^R_{20,20}\non  
    &+&V_0 \big[2 \tilde G^R_{20,20} (\tilde G^R_{00,2-2})^2\non
    &&+\left(\tilde G^R_{2-2,2-2}+\tilde G^R_{2-2,22}\right)
      \left((\tilde G^R_{00,20})^2-\tilde G^R_{00,00} \tilde G^R_{20,20}\right)\big] \ .\non
\ea
For the boost $\vec{P}=(2\pi/L) (1,1,1)$, we obtain:
\begin{align}
\ell&=0,1,& A_1:\hspace{0.4cm} &
(1-V_0\tilde G^R_{00,00})(1-V_1\tilde G^R_{10,10})
\non &&&
-V_0V_1(\tilde G^R_{00,10})^2
 =0\, ,\non
\ell&=1,& E:\hspace{0.4cm} &
1-V_1\tilde G^R_{11,11}=0\, ,\non
\ell&=0,2,& A_1^+:\hspace{0.4cm} &
(1-V_0\tilde G^R_{00,00})(1-V_2\tilde G^R_{20,20})
\non &&&
-V_0V_2(\tilde G^R_{00,20})^2
 =0\, ,\non
\ell&=2,& E^+:\hspace{0.4cm} &
(1-V_2\tilde G^R_{22,22})(1-V_2\tilde G^R_{21,21})
\non &&&
-V_2^2\tilde G^R_{22,2-1}
   \tilde G^R_{2-1,22}=0\, .
\label{rot2}
\end{align}
Finally, we denote that the structures of Eq.~(\ref{gmat}) are the same with rotated $\tilde
G^{R}$, except that the rotation renders many $\tilde G^{R}$ equal to zero, so that the
block-diagonal structure becomes immediately visible.


\subsection{Equations for the determination of the scattering phase shifts}
\label{sec:recon_kappa}

One of the main tasks of this study is the reconstruction of the  $S$-wave phase shift from
lattice data taken in moving frames.  We will assume that suitable interpolating operators are
used in lattice simulations that allow to associate the levels obtained to any of the symmetry
groups discussed. A step forward in this direction has been given, e.g., in
Refs.~\cite{Dudek:2012gj,Gockeler:2012yj} by  constructing a basis of  interpolating operators
transforming  irreducibly under the reduced symmetry of the moving particle in a cubic  box.
Similarly, a complete list of two-particle interpolators is quoted in 
Ref.~\cite{Leskovec:2012gb} for meson momenta $p_i$ where
$p_1,p_2\leq 2\pi\sqrt{3}/L$.

The structure of the equations for the level determination, denoted in the previous section,
lends itself to a suitable strategy: the $P$- or $D$-waves can be reconstructed separately and
serve as input to disentangle the $S$-wave from the representations $A_1$ and $A_1^+$ in which
$P$- or $D$-waves appear mixed with the $S$-wave, respectively. For the one-channel case, a
corresponding set of equations is formulated in the following. 

Note that to disentangle partial waves one needs in principle eigenvalues from different
irreducible representations at exactly the same energy, which will unlikely be the case in an
actual lattice calculation. This requires to make assumptions on the continuity of the amplitude
so that one can interpolate to different energies.  Strategies of how to do this in practice are
discussed in Sec.~\ref{sec:strategies}.

The key point for phase extraction is that, although Eqs.~(\ref{p001sp},
\ref{p001sd}, \ref{rot1}) and (\ref{rot2}) are formulated in terms of the divergent, and thus
cut-off dependent $\tilde G$, the extraction and disentanglement of partial waves can be
formulated entirely in terms of 
\ba
\hat G_{\ell m,\ell'm'}
  &=&\tilde G^R_{\ell m,\ell'm'}
   -\delta_{\ell\ell'}\delta_{mm'}\left(G+\frac{ip}{8\pi E}\right)\non
  &=&\tilde G^R_{\ell m,\ell'm'}-\delta_{\ell\ell'}\delta_{mm'}\,{\rm Re}\,G
\label{ghat}
\ea
with $\tilde G^R$ from Eq.~(\ref{glmrot}) and $G$ from Eq.~(\ref{prop_cont}) (the last line of
Eq.~(\ref{ghat}) is valid above threshold). For the quantity $\hat G$ the dependence on the
cut-off cancels because it depends only on the difference $\tilde G-G$; the quantity $\tilde G$
for the case $\delta_{\ell\ell'}\delta_{mm'}=0$ is convergent anyway: for $\vec q^{\,*}$ much
larger than the considered typical momenta, the sum (c.f. Eq.~(\ref{glmrot})) can be approximated
by the integral and then Eq.~(\ref{norm}) renders the high-momenta contributions to zero.

Below we demonstrate in one example, how to disentangle $S$-waves from the $P$-waves. We
consider the irreducible representation $A_1$. Eqs.~(\ref{p001sp}), (\ref{rot1}) and
(\ref{rot2}) all have the same form in this representation
\be
(1-V_0\tilde G^R_{00,00})(1-V_1\tilde G^R_{10,10})-V_0V_1 (\tilde G^R_{00,10})^2 =0\, .
\label{dette}
\ee
This equation determines the $A_1$ levels for the boosts $\vec P\sim (0,0,1)$,
$(1,1,0)$, $(1,1,1)$. We further introduce
\ba
T_i&=&\frac{V_i}{1-V_i G}=\frac{-8\pi\,E}{p\,\cot\delta_i(p)-i\,p} \ ,\quad\quad
\nonumber\\[2mm]
K_i&=&\frac{V_i}{1-V_i\,{\rm Re}\,G}=\frac{-8\pi\,E}{p\,\cot\delta_i(p)} \ ,\quad\quad i=0,1\, .
\label{phase_connect}
\ea
With these definitions, Eq.~(\ref{dette}) is rewritten as
\be
(1-K_0\hat G_{00,00})(1-K_1\hat G_{10,10})-K_0K_1 (\hat G_{00,10})^2 =0\, . 
\label{dette2}
\ee
Solving this equation with respect to the $S$-wave $\delta_0\equiv \delta_S$ as a function of the
$P$-wave $\delta_1\equiv\delta_P$, we finally obtain 
\be
p\,\cot\delta_0=-8\pi\,E\,\hat G_{00,00}
    +\frac{(8\pi E)^2\,\hat G_{00,10}^2}{p\,\cot\delta_1+8\pi\,E\,\hat G_{10,10}}\, ,
\label{correct}
\ee
which can also be obtained from Eq.~(\ref{dette}) upon substitution of $\tilde G^R$ by $\hat G$
where $\hat G$ is defined in Eq.~(\ref{ghat}). Note that, although in the derivation of
Eq.~(\ref{correct}) we have used the quantities $\tilde G$ and $G$, the final result depends only
on the cut-off independent quantity $\hat G$, as has to be.

If the $P$-wave phase shift vanishes, Eq.~(\ref{correct}) simplifies to
\be
p\,\cot\delta_0=-8\pi\,E\,\hat G_{00,00} \ ,
\label{decouple}
\ee
i.e. the standard L\"uscher formula for pure $S$-wave, in the formulation of
Ref.~\cite{Doring:2011vk}.

Eq.~(\ref{correct}) provides -- in the one-channel case -- the possibility to fully correct for
partial wave mixing at the energy $E$ of the measured level, but it requires the
knowledge of the $P$-wave phase shift $\delta_1$ at that energy. As discussed above,
this knowledge may come from a separate analysis of levels with pure $P$-wave content,
i.e. from other representations than $A_1$, c.f. Eqs.~(\ref{p001sp}), (\ref{rot1}) and
(\ref{rot2}). The following cases allow for this extraction of the $P$-wave:
for $\vec P=(2\pi/L)(0,0,0)$: representation $T_1^-$, c.f. Eq.~(\ref{p000});
for $\vec P=(2\pi/L)(0,0,1)$: representation $E$, c.f. Eqs.~(\ref{p001sp});
for $\vec P=(2\pi/L)(1,1,0)$: representations $B_1$ and $B_2$, c.f. Eq.~(\ref{rot1});
for $\vec P=(2\pi/L)(1,1,1)$: representation $E$, c.f. Eq.~(\ref{rot2}).

Finally, we consider the mixing of $S$- and $D$-waves in case of equal-mass particles.
Extraction of the $D$-waves alone proceeds analogously to the extraction of the $P$-waves in
all cases except  $\vec P=(2\pi/L)(1,1,1)$. In this case, there are two independent solutions
\begin{multline}
p\cot\delta_2=-4\pi E\bigg[\hat G_{22,22}+\hat G_{21,21}\\
\pm
\sqrt{4\hat G_{22,2-1}\hat G_{2-1,22}+\left(\hat G_{22,22}-\hat G_{21,21}\right)^2}\bigg].
\end{multline}
This is related to the fact that, in this case, there are two different representations $E^+$ 
for $\ell=2$.

Finally, the mixing of the $S$- and $D$-waves can be treated analogously to the  mixing of the
$S$- and $P$-waves, since the equations are always linear in $\cot\delta_0$. In the case $\vec
P=(2\pi/L)(1,1,0)$ the final equations are a bit cumbersome and contain terms quadratic in
$\cot\delta_2$ (not explicitly quoted here).

For the two-channel case, the structure of the determinant (\ref{det0}) is more complicated,
because the $V$ and $\tilde G$ are matrices in channel space as discussed in
Sec.~\ref{sec:combi}. There is not much point to formulate analytic formulae as in this
section, because for a disentanglement of the $S$-wave in a channel, one needs to know not only
the $P$- or $D$-wave phase shift, but also inelasticities and phase shifts in the other
channel. For $S$-wave and $\vec P=\vec 0$, the reconstruction of the amplitude in the
two-channel case has been discussed and solved in Ref.~\cite{Doring:2011vk}.  In this study, we
will construct synthetic data using the full two-channel case and reconstruct the one-channel
phase shifts using the framework developed in this section. Below the inelastic threshold, this
provides a very good approximation whose validity will be discussed.


\section{The $\boldsymbol{\kappa}\mathbf{(800)/K^*(892)}$ system}
\label{sec:kappa}


\subsection{Level spectrum}
\label{sec:spectrum}


To study the mixing of partial waves and apply the framework developed in previous sections,
the case of coupled-channel $\pi K,\,\eta K$ scattering in a moving frame is considered. In
isospin $I=1/2$, strangeness $S=-1$, we take account of the $S$-wave with the $\kappa(800)$
resonance, the $P$-wave with the $K^*(892)$, and neglect all higher partial waves.
$S$-wave $\pi K$ scattering and the $\kappa$ resonance 
have been addressed in recent lattice calculations, e.g. in Refs.~\cite{Fu:2011xw,Lang:2012sv}.

\begin{figure}
\begin{center}
\hspace*{0.15cm}\includegraphics[width=0.4625\textwidth]{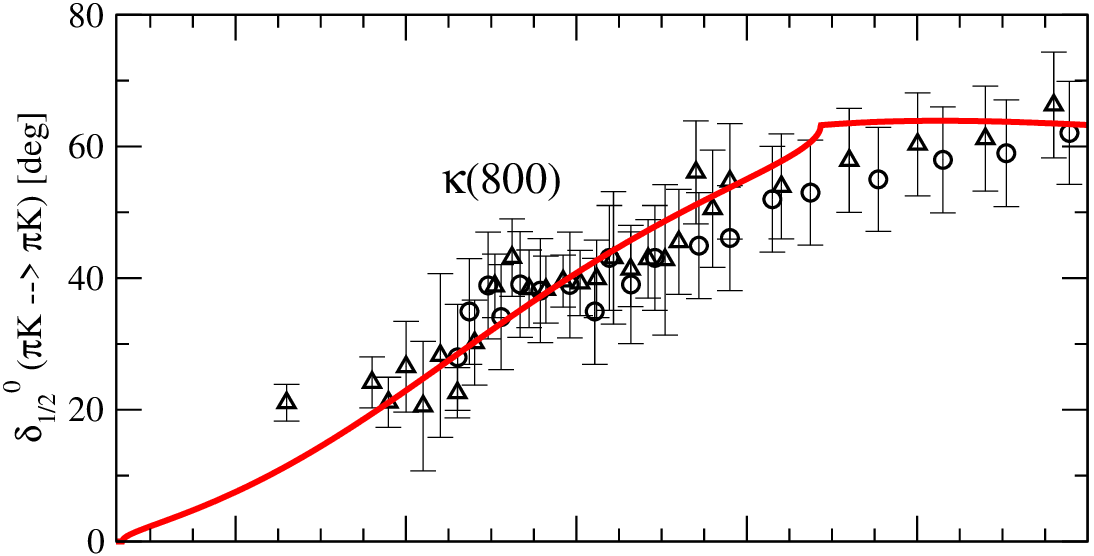} \\
\vspace*{-0.155cm}
\includegraphics[width=0.47\textwidth]{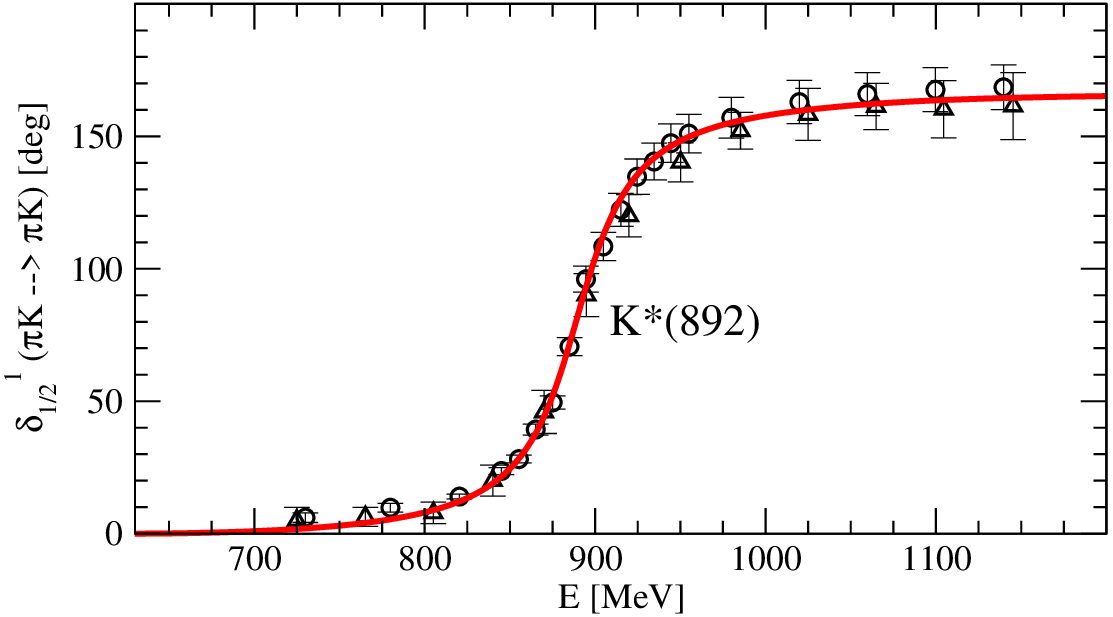} 
\end{center}
\caption{Solution for the isospin $I=1/2$, strangeness $S=-1$ meson-meson interaction (solid
lines) of Ref.~\cite{Doring:2011nd}. Partial wave data for $\delta_{1/2}^0$:
circles~\cite{Aston:1987ir}, triangles up: average as defined in ref.~\cite{Oller:1998zr}. For
$\delta_{1/2}^1$: triangles up~\cite{Mercer:1971kn}, circles~\cite{Estabrooks:1977xe}
}
\label{fig:phase_shifts}
\end{figure}

For the hadronic amplitude in the infinite volume, we use the solution obtained in
Ref.~\cite{Doring:2011nd} by fitting the low energy constants $L_1$ to $L_8$ to strangeness
$S=0,-1$ partial wave data in $S$- and $P$-waves. That solution is obtained using the inverse
amplitude method of Ref.~\cite{Oller:1998hw}. In Eq.~(\ref{replaceia2}) the connection to the
present framework is quoted. The phase shifts are shown in Fig.~\ref{fig:phase_shifts}.
For the $S$-wave we observe a pronounced effect due to the $\eta K$ threshold which shows that
this channel can have noticeable influence. For the $P$-wave, the $\eta K$ channel plays almost
no role.

\begin{figure}
\begin{center}
\includegraphics[width=0.49\textwidth]{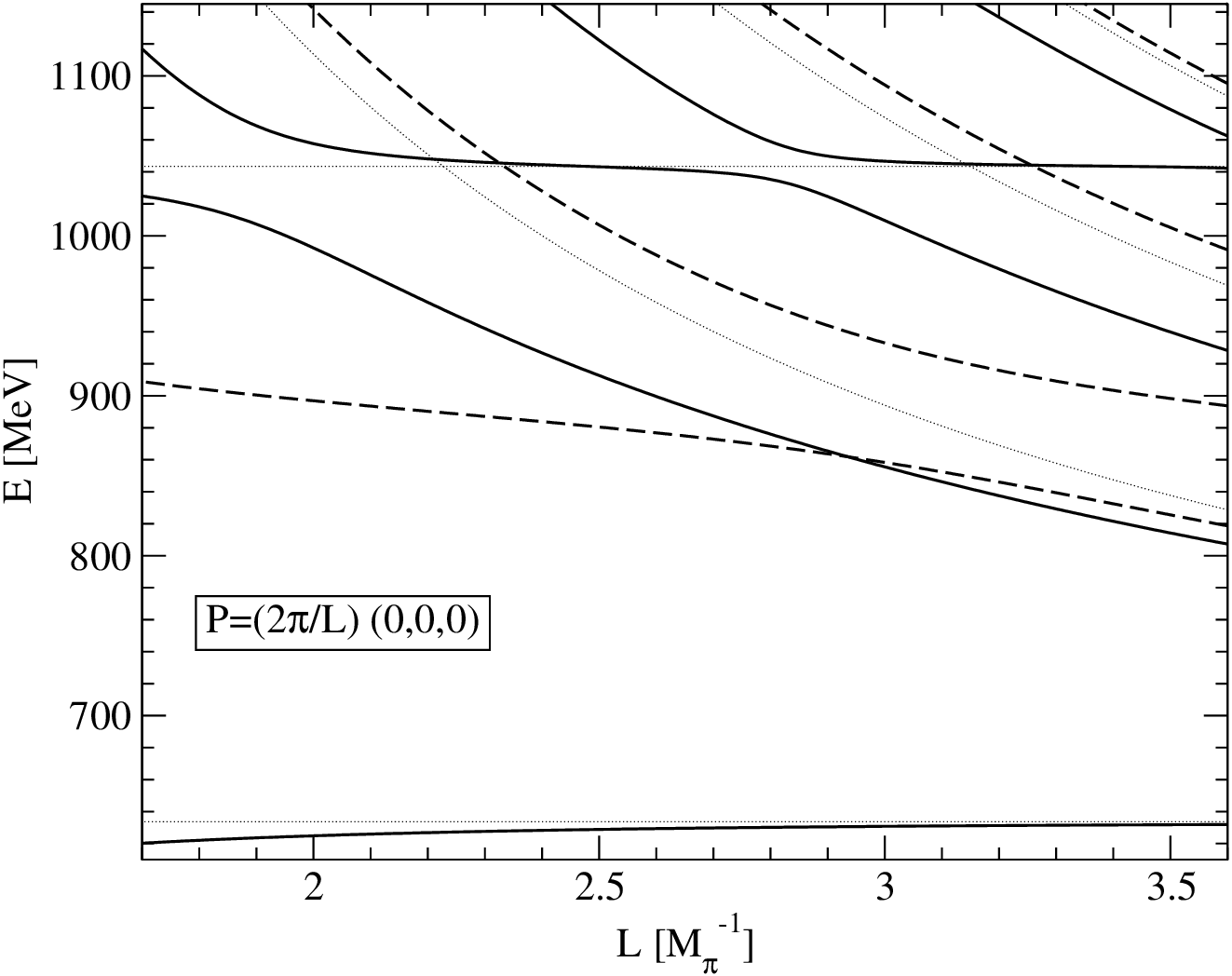}
\end{center}
\caption{Spectrum of the $S$- and $P$-wave system in coupled-channel scattering ($\pi K$, $\eta
K$) for $\vec P=(0,0,0)$. Solid (dashed) lines: $S$-wave ($P$-wave). The fine dotted lines show
the non-interacting levels.}
\label{fig:levels_000_overview}
\end{figure}

Using the equations of Sec.~\ref{akakisolve} or \ref{michasolve}, the lattice spectrum for the
different boosts can be predicted as a function of the box length $L$. For $\vec P=\vec 0$, the
result is shown in Fig.~\ref{fig:levels_000_overview}. Here, $S$- and $P$-waves do not mix.
We observe an $S$-wave level close to the $\pi K$ threshold as well as avoided level crossing at
the $\eta K$ threshold for the $S$-wave. These features have been discussed extensively in
Refs.~\cite{Doring:2011vk,Doring:2011nd}. In particular, the avoided level crossing is a signal
of the $\eta K$ threshold (c.f. fine dashed horizontal line) and not of the $\kappa(800)$. 

\begin{figure}
\begin{center}
\includegraphics[width=0.49\textwidth]{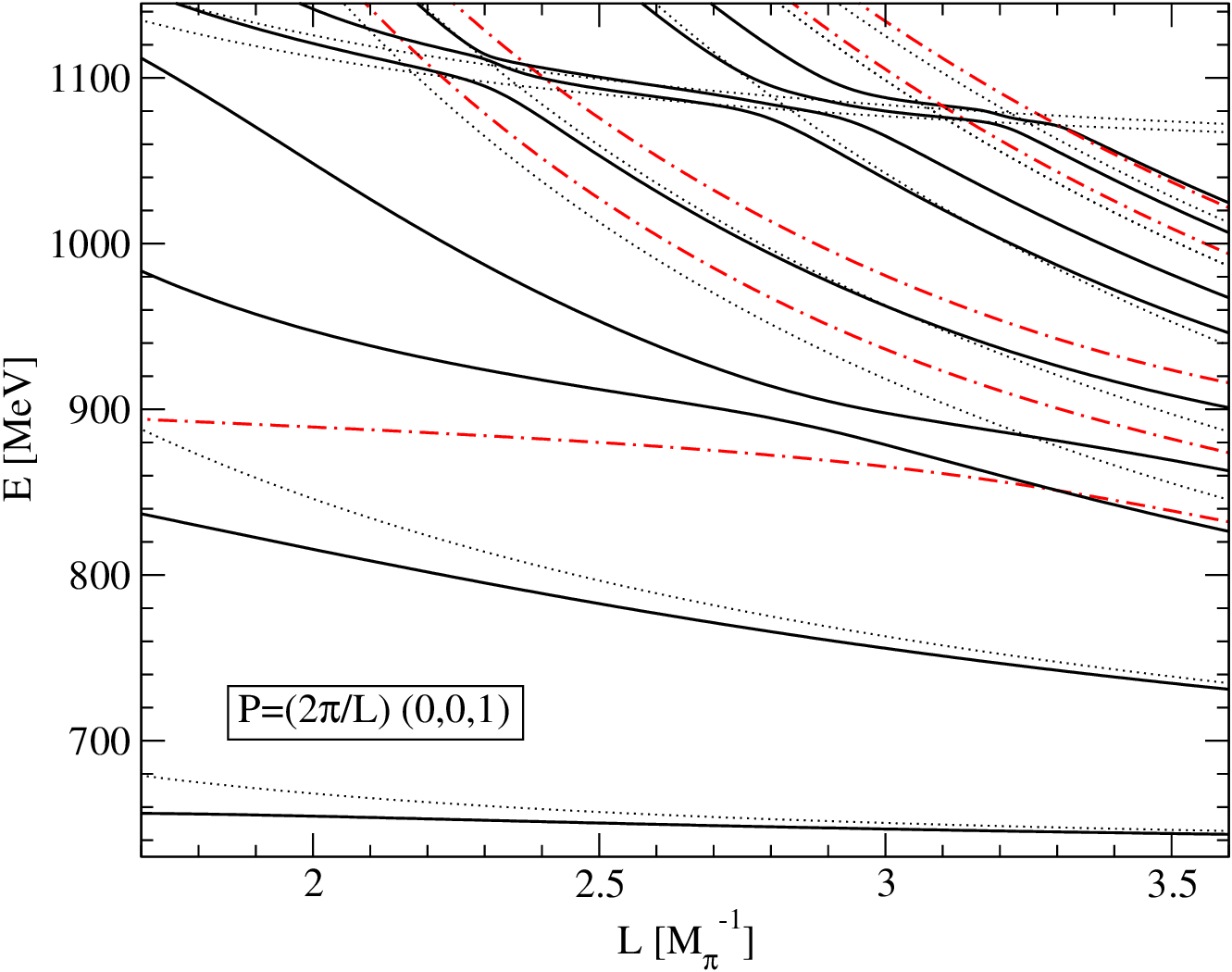}
\end{center}
\caption{Spectrum of the $S$- and $P$-wave system in coupled-channel scattering ($\pi K$, $\eta
K$) for $\vec P=(2\pi/L)(0,0,1)$. Solid lines: levels with $SP$-wave mixing from the
irreducible representation $A_1$. Dash-dotted (red) lines: levels from only $P$-wave,
irreducible representation $E$. The fine dotted lines show the boosted non-interacting levels.}
\label{fig:levels_001_overview}
\end{figure}

\begin{figure}
\begin{center}
\includegraphics[width=0.49\textwidth]{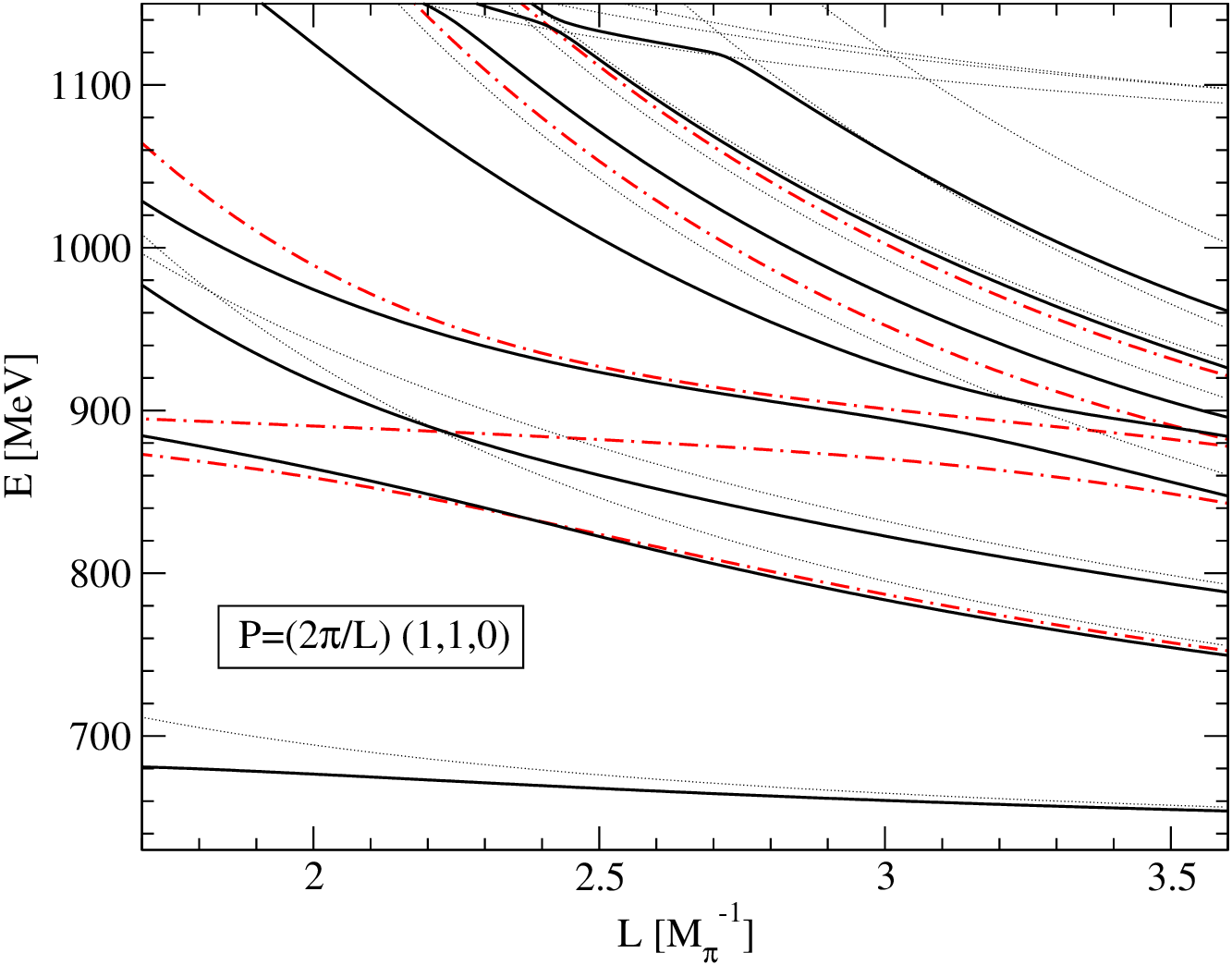}
\end{center}
\caption{Spectrum of the $S$- and $P$-wave system in coupled-channel scattering ($\pi K$, $\eta
K$) for $\vec P=(2\pi/L)(1,1,0)$. Solid lines: levels with $SP$-wave mixing from the irreducible
representation $A_1$. Dash-dotted (red) lines: levels from only $P$-wave, irreducible
representations $B_1$ and $B_2$. The fine dotted lines show the boosted non-interacting levels.}
\label{fig:levels_011_overview}
\end{figure}

\begin{figure}
\begin{center}
\includegraphics[width=0.46\textwidth]{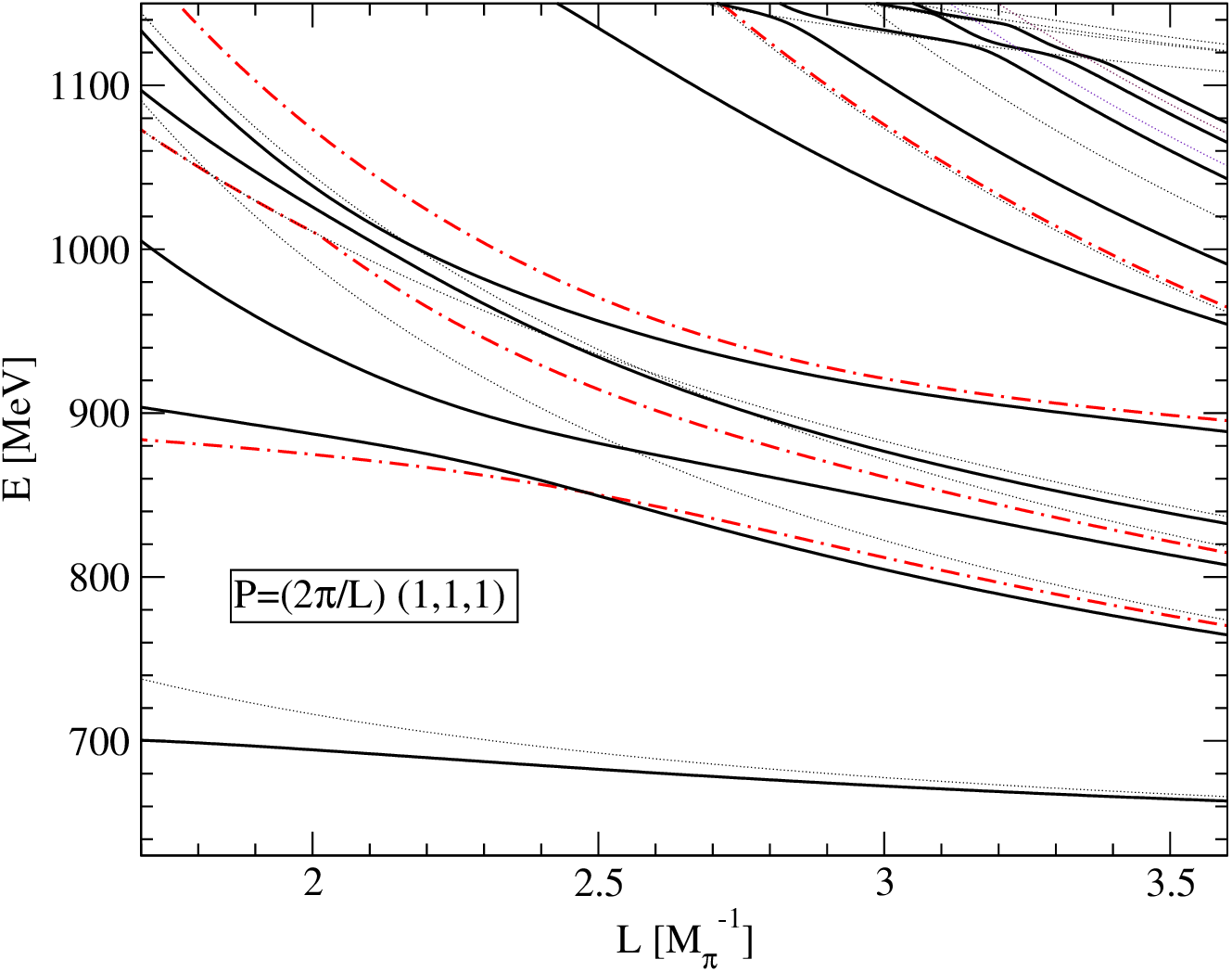}
\end{center}
\caption{Spectrum of the $S$- and $P$-wave system in coupled-channel scattering ($\pi K$, $\eta
K$) for $\vec P=(2\pi/L)(1,1,1)$. Solid lines: levels with $SP$-wave mixing from the
irreducible representation $A_1$. Dash-dotted (red) lines: levels from only $P$-wave,
irreducible representation $E$. The fine dotted lines show the boosted non-interacting levels.}
\label{fig:levels_111_overview}
\end{figure}

As we choose a finite boost $\vec P\neq\vec 0$, the level spectrum becomes more complex. For
$P=(2\pi/L)(0,0,1)$, $P=(2\pi/L)(1,1,0)$, and $P=(2\pi/L)(1,1,1)$ the resulting levels as a
function of $L$ are shown in Figs.~\ref{fig:levels_001_overview},
\ref{fig:levels_011_overview}, and \ref{fig:levels_111_overview}, respectively.

As the boost increases, the (non-interacting) levels are in general moved upwards in energy.
Close to the lowest boosted non-interacting level, i.e. the boosted $\pi K$ threshold, one always
finds  a  level in which $S$- and $P$-waves are mixed (solid lines, representation
$A_1$)~\footnote{The levels of the representation $A_1$ are determined from Eq.~(\ref{dette}).
The first term of  Eq.~(\ref{dette}) suggests the occurrence of two levels that undergo a mixing
through the second term $\sim V_0V_1$. However,  as e.g. Fig.~\ref{fig:levels_001_overview}
shows, there is only one level close to the first non-interacting energy, and only one close to
the second non-interacting energy. Indeed, it can be shown that this must be the case, by using
the known angular structure of the $\hat G_{\ell m,\ell'm'}$.}. At higher energies, apart from
levels of this kind, also levels from other representations start to appear containing only
$P$-wave without $S$-wave admixture and allowing for a separate analysis of the $P$-wave.


\subsection{Leading behavior of the level shifts}
\begin{figure}
\begin{center}
\includegraphics[width=0.49\textwidth]{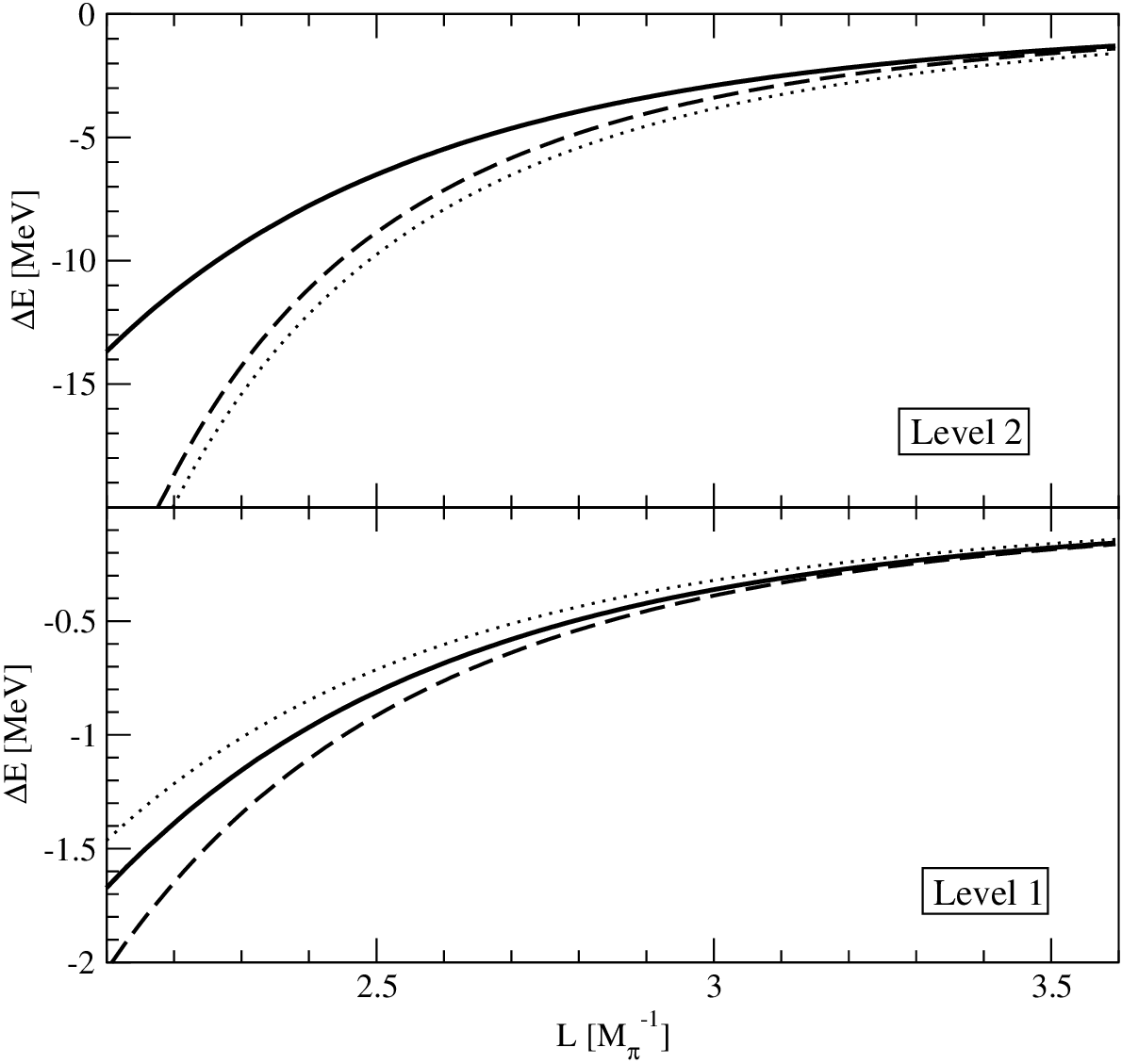}
\end{center}
\caption{Level shift through partial wave mixing for $\vec P=(2\pi/L)(0,0,1)$. Below: for the
lowest level from Fig.~\ref{fig:levels_001_overview}. Above: for the following level. The solid
lines show the actual level shift,  the dashed lines the result of Eq.~(\ref{new_shiftt}) and
the dotted lines the leading behavior from Eq.~(\ref{shiftt}).}
\label{comparede}
\end{figure}

The levels from the $A_1$ representation get shifted once the mixing is taken into account. To
understand this effect, we consider mixing in the limit of small phase shifts.  In the absence
of a $P$-wave ($V_1=0$), Eq.~(\ref{dette}) reduces to 
\be
1-V_S\tilde G^R_{00,00}(E=E_S)=0
\ee
which determines the position $E_S$. The level shift is
\be
\Delta E=E_{SP}-E_S
\ee
with $E_{SP}$ the position of the level with mixing, given by the solution of
Eq.~(\ref{dette}). The shift can be approximately calculated by Taylor expanding
Eq.~(\ref{dette}) around $E=E_S$. It is straightforward to show that then approximately
\be
\Delta E\simeq\delta_P\,\frac{8\pi E_S}{p}\,
\frac{(\hat G^R_{00,10})^2}{\partial\,\hat G^R_{00,00}/\partial E}
\label{new_shiftt}
\ee
where $\delta_P$ is the $P$-wave phase shift, $p$ the three-momentum of the $\pi$ and $K$ in
the two-particle rest frame, and $\hat G$ defined in Eq.~(\ref{ghat}). The actual shift $\Delta
E$ is shown with the solid lines in Fig.~\ref{comparede}, the result from
Eq.~(\ref{new_shiftt}) is indicated with the dashed lines.
Indeed, for the lowest level where $\delta_1$ is small Eq.~(\ref{new_shiftt}) provides a good
approximation.

To understand the structure of the shift, we can expand $\hat G$ around the pole at $E=E_0$
where $E_0$ is the boosted non-interacting energy,
\ba
\hat G^R_{\ell m,\ell' m'}&=&\frac{1}{L^3}
\sum_{\vec n}h^{(\vec n)}_{\ell m,\ell' m'}\,I(q^*)\non
&=&\frac{a_{-1}\,h^{(\vec n=\vec 0)}_{\ell m,\ell' m'}}{E-E_0}+R_{\ell m,\ell' m'}
\label{expagtil}
\ea
where $h^{(\vec n)}_{\ell m,\ell' m'}$ contains the angular structure,  $a_{-1}\,h^{(\vec
n=\vec 0)}_{\ell m,\ell' m'}$ is the residue, and $R_{\ell m,\ell' m'}$ is  regular in the
vicinity of $E_0$. The residue is readily evaluated,
\be
a_{-1}=\frac{1}{L^3}\,\frac{1}{2\omega_1\omega_2}\frac{\omega_1+\omega_2}{E+\omega_1+\omega_2}
\ee
where the energies $\omega_i$ are evaluated with the boosted vectors $\vec q^{\,*}\equiv\vec
q^{\,*}(\vec q=\frac{2\pi}{L}(0,0,0))$ for the lowest level and  $\vec q^{\,*}(\vec
q=\frac{2\pi}{L}(0,0,1))$ for the following one. Using $\omega_1+\omega_2\simeq E$ and
substituting Eq.~(\ref{expagtil}) in Eq.~(\ref{new_shiftt}) we obtain
\be
\Delta E\simeq -\frac{6\pi E_S\,\delta_P}{L^3\,p\,\omega_1\omega_2}
\label{shiftt}
\ee
for the level shift. The result of this approximation is shown with the dotted lines in
Fig.~\ref{comparede}. This equation shows that the shift is to leading order in $E$
proportional to $\delta_P$ and to $1/L^{-3}$. This $L^{-3}$ behavior is similar to the one of
the scattering length~\cite{Luscher:1986pf,Doring:2011nd}. While Eq.~(\ref{shiftt}) is useful to
understand the qualitative behavior of the level shift, its quantitative use is limited.


\subsection{Disentangling the $\boldsymbol{\kappa}\mathbf{(800)/K^*(892)}$ system}

The $S$-wave can be extracted from lattice data of the representations $A_1^+$ (boost $\vec
P=\vec 0$) and $A_1$ (boosts $\vec P=(2\pi/L)(0,0,1)$, $(2\pi/L)(0,0,1)$, $(2\pi/L)(1,1,0)$,
$(2\pi/L)(1,1,1)$). While for $\vec P=\vec 0$ the $S$-wave does not mix with the $P$-wave, for
the higher boosts mixing occurs in the $A_1$ representation, c.f. Eqs.~(\ref{p000},
\ref{p001sp}, \ref{rot1}, \ref{rot2}). We have derived Eq.~(\ref{correct}) that allows to
disentangle the $S$-wave for the one-channel problem provided that the $P$-wave is known from
an analysis of other levels, as discussed in Sec.~\ref{sec:recon_kappa}.

In this section, we apply Eq.~(\ref{correct}) to the levels corresponding to $A_1$ shown in
Figs.~\ref{fig:levels_001_overview} to \ref{fig:levels_111_overview}. It should be stressed
that those levels have been obtained with the full two-channel formalism as described in
Sec.~\ref{sec:combi}, based on the global fit of low energy constants to $S$- and $P$-wave
partial wave data as described in Sec.~\ref{sec:spectrum}. In contrast, we will extract the
phase shift from these levels using the one-channel equation (\ref{correct}). Below the
inelastic threshold, given by the $\eta K$ channel, this is expected to be a good
approximation. Indeed, in Refs.~\cite{Bernard:2010fp,Doring:2011vk}, the so-called
pseudo-phase, i.e. the phase extracted with a one-channel formalism from a two-channel problem,
provides an excellent approximation to the actual phase up to energies close below the
inelastic threshold. 

\label{sec:disentangle_kappa}
\begin{figure}
\begin{center}
\includegraphics[width=0.47\textwidth]{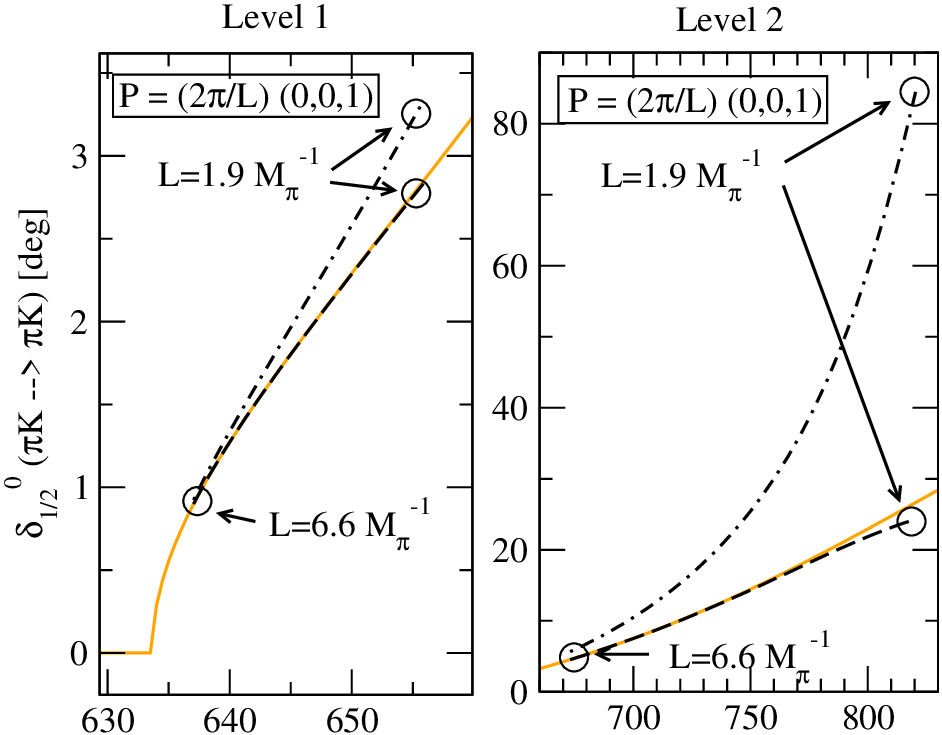} \\
\includegraphics[width=0.47\textwidth]{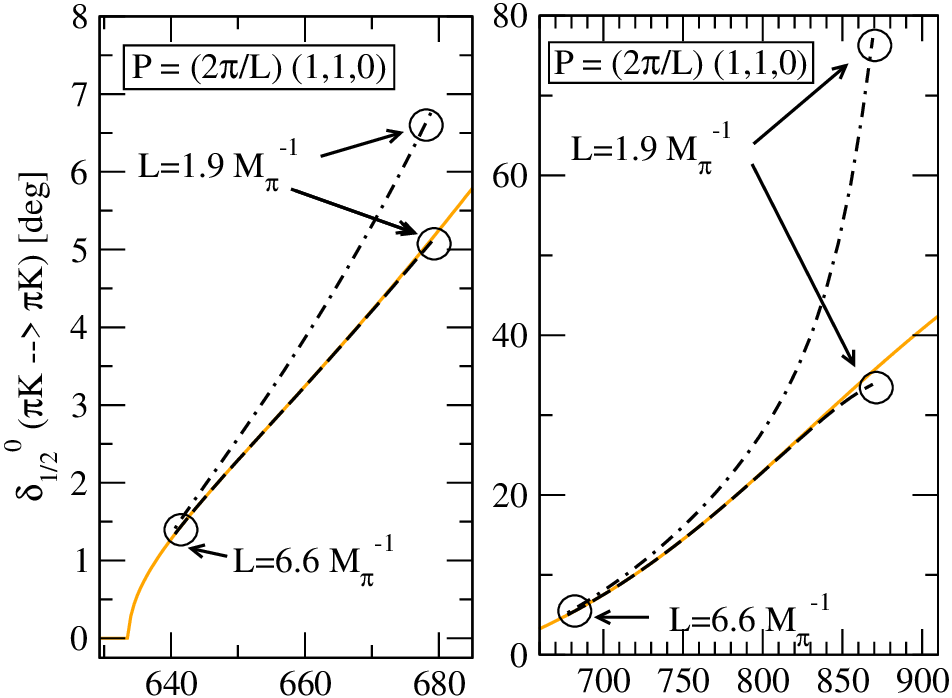} \\
\includegraphics[width=0.47\textwidth]{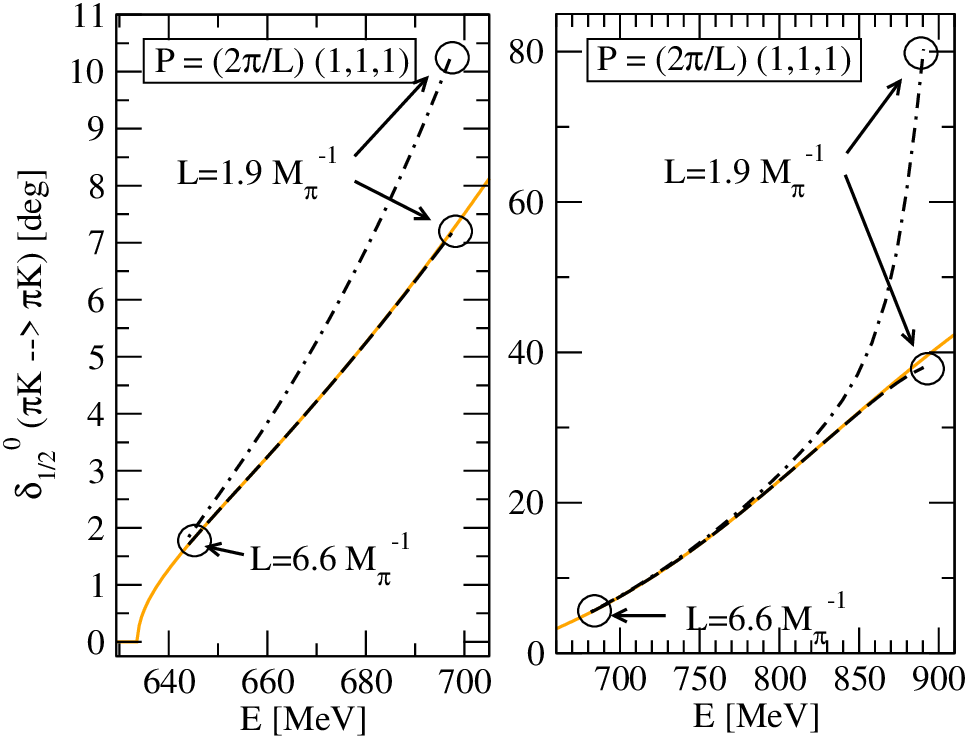} 
\end{center}
\caption{Actual phase shifts [solid (orange) lines], identical to the result shown in
Fig.~\ref{fig:phase_shifts} and the extraction from the first and second level, using
Eq.~(\ref{correct}) for the disentanglement (dashed lines) or Eq.~(\ref{decouple}) in
which the partial wave mixing is neglected (dash-dotted lines). Results are shown for the first
three boosts $\sim(0,0,1),\, (1,1,0)$, and $(1,1,1)$. In the figure, also the corresponding
values of the box size $L$ are indicated.}
\label{fig:34vs36}
\end{figure}

In Fig.~\ref{fig:34vs36}, the actual $\pi K$ $S$-wave phase shift, from the full coupled-channel
system, is shown with the solid (orange) lines, identical to the  corresponding curve in
Fig.~\ref{fig:phase_shifts}. Consider the first two $A_1$ levels, tied to the first three boosts
[solid (black) lines in Figs.~\ref{fig:levels_001_overview}, \ref{fig:levels_011_overview}, and
\ref{fig:levels_111_overview}]. In those levels, the $S$-wave mixes with the $P$-wave, and
Eq.~(\ref{correct}) provides the possibility to disentangle the $S$-wave by using the known
$P$-wave shown in Fig.~\ref{fig:phase_shifts}. The result is shown with the long-dashed lines in
Fig.~\ref{fig:34vs36}. In the figure, we also indicate to which box size $L$ the extracted phase
shift corresponds (c.f. again Figs.~\ref{fig:levels_001_overview}, \ref{fig:levels_011_overview},
and \ref{fig:levels_111_overview} to see the connection of energies and box size, given by the
levels).  As a test of the formalism, we have shown that once levels are generated from the
hadronic model in a reduction to one channel, Eq.~(\ref{correct}) ensures the exact
reconstruction of the phase, as must be.

In general, the agreement with the original phase shift is excellent. Only at higher energies
there are small deviations, coming from the more and more important $\eta K$ channel. It should
be noted that the effect on the reconstruction of the phase is small but the reconstruction of
the pole position might be affected. This has been shown in Ref.~\cite{Doring:2011nd} for the
$\kappa(800)$ and $\vec P=\vec 0$.

Second, we quantify the effect if partial wave mixing is neglected (in the one-channel
extraction scheme we are are using). In that case, Eq.~(\ref{correct}) reduces to
Eq.~(\ref{decouple}) which is the ordinary L\"uscher equation for a boosted $S$-wave system. The
results are shown with the dash-dotted lines in Fig.~\ref{fig:34vs36}. For very large $L$, we
indeed observe that the extracted and the actual phase are similar. In other words, the $P$-wave
decouples from the $S$-wave in the infinite volume limit as must be. However, even for
$L>3M_\pi^{-1}$ we already observe large deviations and for smaller box sizes than $L\sim 2
M_\pi^{-1}$, a reliable extraction of the phase, let alone the pole position of the
$\kappa(800)$, is not possible any more. 

In summary, it is crucial to take effects from partial wave mixing into account while effects
from inelastic channels can be neglected to a good accuracy as long as one considers energies
only below the inelastic threshold. As a side remark, we would like to stress that in different
physical contexts it might be crucial to use a two-channel extraction scheme, in particular if
thresholds are close to each other as in case of the
$\Lambda(1405)$~\cite{Lage:2009zv,Doring:2011ip,MartinezTorres:2012yi}, or if one wants to
extract phases and poles close to thresholds as in case of the $f_0(980)$~\cite{Doring:2011vk}.


\section{The $\boldsymbol{\sigma(600)}$ in a moving frame}
\label{sec:sigma}
\subsection{Level spectrum and resonance extraction}
\label{sec:spectrumsigma}
\begin{figure}
\begin{center}
\includegraphics[width=0.47\textwidth]{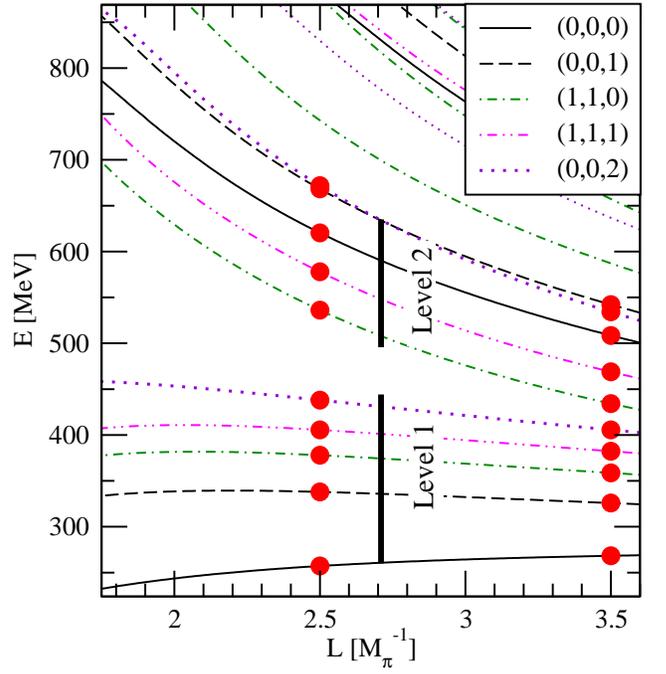} 
\end{center}
\caption{Levels of the $(I=L=S=0)$ meson-meson system [$\sigma(600)$ and $f_0(980)$] for
different boosts. The symbols at $L=2.5M_\pi^{-1}$ and $L=3.5M_\pi^{-1}$ indicate the 20
synthetic data points taken for the reconstruction of phase and $\sigma(600)$ pole position
[results in Figs.~\ref{fig:phases_sigma} and \ref{fig:poles_sigma}].  }
\label{fig:levels_sigma}
\end{figure}
The first few levels for the $I=L=S=0$ quantum numbers are shown in Fig.~\ref{fig:levels_sigma}
for the first five boosts. To obtain these levels, the $S$-wave two-channel potential
$(\pi\pi,\,\bar KK)$ from the inverse amplitude method, $V_S^{(ij)}\equiv V^{\rm IAM}$, given
in Eq.~(\ref{replaceia2}), is used in Eq.~(\ref{det0}) to determine the levels. For $V^{\rm
IAM}$, the corresponding solution of Ref.~\cite{Doring:2011nd} for the underlying hadronic
interaction is used.

\begin{figure}
\begin{center}
\includegraphics[width=0.47\textwidth]{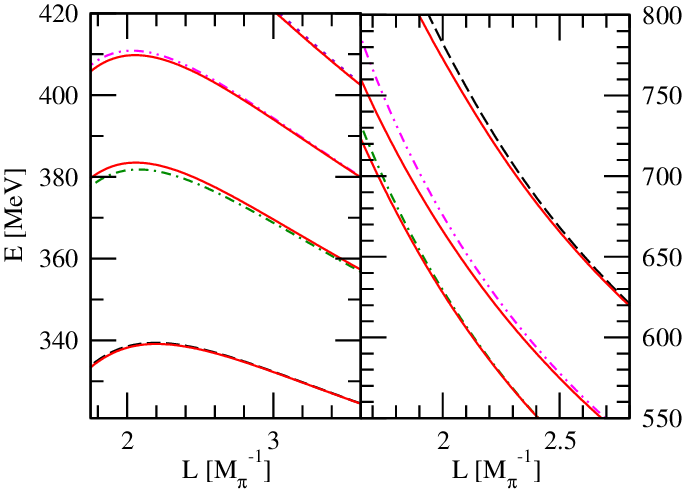} 
\end{center}
\caption{Levels from the $A_1^+$ representation as in Fig.~\ref{fig:levels_sigma}. In addition,
the (red) solid lines show the levels mixed with $D$-wave.
}
\label{fig:levels_sigma_mixing}
\end{figure}

To quantify the expected error from partial wave mixing, we have calculated the level shift due
to the $S$-wave mixing with the small isospin zero $D$-wave. 
Note that for the considered equal-mass case ($\pi\pi$ and $\bar KK$), there is no
$SP$-wave mixing even for $\vec P\neq 0$. Anyway, the $\pi\pi$ system in $P$-wave has isospin one.
For the $SD$-wave mixing, the phenomenological
parameterization of the $D$-wave of Ref.~\cite{GarciaMartin:2011cn} is used that
serves to construct a $D$-wave to $D$-wave transition $V_D^{(2,2)}$ as described in
Appendix~\ref{appe}. We do not include a $\bar KK$ channel in $D$-wave as there is no
phenomenological reason for it and we are far below the $\bar KK$ threshold, anyway. We have,
for the $A_1^+$ representation, the levels given by Eq.~(\ref{det0}) where 
\ba
V&=&
\begin{pmatrix}
V_S^{(11)} 	& 	V_S^{(12)}	& 0   \\
V_S^{(21)} 	& 	V_S^{(22)}	& 0   \\
0		&	0		& V_D^{(22)}
\end{pmatrix}\non
\tilde G&=&
\begin{pmatrix}
\tilde G_{00,00}^{R\,(1)}	&	0			&	0     			\\
0				& \tilde G_{00,00}^{R\,(2)}	& \tilde G_{00,20}^{R\,(2)} 	\\[1mm]
0				& \tilde G_{20,00}^{R\,(2)}	& \tilde G_{20,20}^{R\,(2)} 
\end{pmatrix}
\ea 
for the boosts $\vec P=(2\pi/L)(0,0,1),$ $(2\pi/L)(1,1,1)$, and $(2\pi/L)(0,0,2)$. For the boost $\vec P=(2\pi/L)(1,1,0)$,
\ba
V&=&
\begin{pmatrix}
V_S^{(11)} 	& 	V_S^{(12)}	& 0   		& 0		& 0 \\
V_S^{(21)} 	& 	V_S^{(22)}	& 0   		& 0		& 0 \\
0		&	0		& V_D^{(22)}	& 0		& 0 \\
0		&	0		& 0		& V_D^{(22)}	& 0 \\
0		&	0		& 0		& 0		& V_D^{(22)} \\
\end{pmatrix}\non
\tilde G&=&
\begin{pmatrix}
\tilde G_{00,00}^{R\,(1)}	&	0			&	0     			&	0			&	0     		\\[1mm]
0				& \tilde G_{00,00}^{R\,(2)}	& \tilde G_{00,2-2}^{R\,(2)} 	& \tilde G_{00,20}^{R\,(2)}	& \tilde G_{00,22}^{R\,(2)}\\[1.4mm]
0				& \tilde G_{2-2,00}^{R\,(2)}	& \tilde G_{2-2,2-2}^{R\,(2)} 	& \tilde G_{2-2,20}^{R\,(2)}	& \tilde G_{2-2,22}^{R\,(2)}\\[1.4mm]
0				& \tilde G_{20,00}^{R\,(2)}	& \tilde G_{20,2-2}^{R\,(2)} 	& \tilde G_{20,20}^{R\,(2)}	& \tilde G_{20,22}^{R\,(2)}\\[1.4mm]
0				& \tilde G_{22,00}^{R\,(2)}	& \tilde G_{22,2-2}^{R\,(2)} 	& \tilde G_{22,20}^{R\,(2)}	& \tilde G_{22,22}^{R\,(2)}
\end{pmatrix}
\ea 
where the index $(1)$ labels the $\bar KK$ channel and $(2)$ the $\pi\pi$ channel. The
resulting $SD$-wave mixed levels of the $A_1^+$ representation are shown with the (red) solid
lines in Fig.~\ref{fig:levels_sigma_mixing}, together with the $S$-wave levels without mixing
from Fig.~\ref{fig:levels_sigma}.
As the figure shows, the level shift from mixing, $\Delta E$ can reach up to $10$~MeV at
$L=2M_\pi^{-1}$, but stays below $3$~MeV for $L=2.5M_\pi^{-1}$ and larger. The shift $\Delta E$
increases not only for smaller $L$, but also for higher energies where $\delta_D$ is larger. 
Eq.~(\ref{shiftt}) indeed shows that this is the expected behavior.

Note that the level shift from the $SD$-mixing is much smaller than the one of the $\pi
K,\,\eta K$ system discussed in Sec.~\ref{sec:kappa}, because the $\pi\pi$ $D$-wave is much
smaller than the $\pi K$ $P$-wave in which the $K^*(892)$ resides.


\subsection{Extraction of the $\boldsymbol{\sigma(600)}$}
\label{sec:extract}

As the level shift from mixing is smaller than $3$~MeV for $L=2.5M_\pi^{-1}$ and larger, we
concentrate on these box sizes and can neglect the mixing; for the reconstruction of the
$\sigma(600)$ we will assume 10~MeV errors on the synthetic data so that this assumption is safe.
In this section, we concentrate on the extraction of the $\sigma(600)$ resonance and do not
analyze the $f_0(980)$. The latter resonance would require a two-channel extraction scheme that
has already been discussed in Ref.~\cite{Doring:2011vk}, and the extension to moving frames is
in principle straightforward. 

As Fig.~\ref{fig:levels_sigma} shows, with larger boosts the first and second level from the
$A_1^+$ representation move towards higher energies. In this way one can cover the entire
energy region from threshold up to $E=700$~MeV with the boosted first two levels. This
demonstrates the advantage lying in the use of moving frames: in the conventional L\"uscher
approach, $L$ is varied at $\vec P=\vec 0$. This means that, first, many different lattice
setups have to be calculated. Second, for the box sizes considered $2<L<3.6M_\pi^{-1}$, the
energy region around $E\sim 400-500$~MeV, i.e. precisely where the real part of the pole
position of the $\sigma(600)$ is located~\cite{Caprini:2003ta}, is not covered by any level,
making the extraction of the $\sigma(600)$ generically more difficult, as has also been noted
in Ref.~\cite{Doring:2011nd}.

In the absence of actual lattice data, we generate 20 synthetic data points from the levels of
Fig.~\ref{fig:levels_sigma} at two values of $L$, $L=2.5M_\pi^{-1}$ and $L=3.5M_\pi^{-1}$, as
indicated with the dots in the figure. A 10~MeV error is assigned to each data point. With
such-defined data, the task is to reconstruct the $\pi\pi$ phase shifts and the $\sigma(600)$
pole. 

For this, we use the parameterization of the one-channel potential from
Ref.~\cite{Doring:2011nd},
\ba
V^{\rm fit}&=&\left(\frac{V_2-V_4^{\rm fit}}{V_2^2}\right)^{-1},\non 
V_4^{\rm fit}&=&a+b(s-s_0)+c(s-s_0)^2+d(s-s_0)^3
\label{vfit}
\ea
with $V_2\equiv V_{\rm LO}$ the fixed LO term of the chiral expansion.  In other words, we take
the form of the inverse amplitude (c.f. Eq.~(\ref{replaceia2})), leaving the LO term $V_2$ as
given by chiral symmetry, and expand $V_4$ in powers of $s$. As expansion point we choose
$s_0=(400\,{\rm MeV})^2$.

The choice of this potential and its advantages have been extensively discussed in
Ref.~\cite{Doring:2011nd}. We denote here that the explicit inclusion of the model-independent,
well-known lowest order term $V_2$ greatly helps stabilizing the extraction. The higher powers
of $s$ account for corrections from the next-to-leading and higher order terms. Note that the
polynomial NLO contributions~\cite{Oller:1998hw} can be approximately taken account of, because
the Mandelstam variable $t$ and $u$ can be expanded in $s$. The effect from the $\bar KK$
channel and its branch point at $E=2M_K$, as well as the left-hand cut, is also absorbed in
this expansion. 

One could in principle include these non-analyticities explicitly in the fit potential.
However, they are not well fixed because they lie much higher or much lower in energy. Given
lattice data in a relatively narrow window in energy, no improvement is expected but instead
large correlations of the corresponding new parameters will arise. The expansion of these
effects in a power series in $s$, as provided in Eq.~(\ref{vfit}), allows for a systematic
improvement and provides a set of parameters with relatively small correlations.

One should bear in mind, though, that this is an approximative procedure. Other than in
Ref.~\cite{Doring:2011vk}, where the analytic form of the fit potential comprised the assumed
hadronic interaction -- this was possible because the model interaction was from lowest order
only -- the potential from Eq.~(\ref{vfit}) can only approximatively absorb the discussed
effects. The strategy is then to perform different fits with increasing powers of $s$ until
convergence is observed as discussed in Ref.~\cite{Doring:2011nd} in detail.


\subsubsection{Extraction strategies with partial wave mixing} 
\label{sec:strategies}
For the $\pi\pi$ $S$-wave
sufficiently below the $\bar KK$ threshold, we can neglect the partial wave mixing as discussed
before, but we give an outlook how to proceed in case it cannot be neglected as for the
$\kappa/K^*$ system discussed in Sec.~\ref{sec:kappa}. On one hand, a separate extraction of $P$-
or $D$-waves is possible from the structure of the irreducible representations $B_1$, $B_2$, $E$
($P$-wave) and $A_2^+$, $B_1^+$, $B_2^+$, $E^+$ ($D$-wave) as discussed following
Eq.~(\ref{decouple}), see also Ref.~\cite{Leskovec:2012gb} (we assume there is no mixing between
$P$- and $D$-wave). However, those phase shifts can only be extracted at  scattering energies
different from where the $A_1^{(+)}$ levels are situated, from which the $S$-wave can be
reconstructed via Eq.~(\ref{correct}) in the one-channel case. For reference, see
Figs.~\ref{fig:levels_001_overview} to \ref{fig:levels_111_overview}. It is then necessary to
make some minimal assumptions on the $P$- or $D$-wave phase shifts, such that the underlying
potential can be expanded in energy as discussed in Sec.~\ref{sec:extract}, c.f.
Eq.~(\ref{vfit}). This has also been recognized in Ref.~\cite{Leskovec:2012gb}.

Instead of determining first the higher partial wave and then the $S$-wave, it might be
advantageous to simultaneously fit the different levels with $V_S$ and $V_P$ (or $V_S$ and
$V_D$), both of them expanded in energy as in Eq.~(\ref{vfit}). Such a procedure would be in
analogy with the two-channel extraction scheme developed in Ref.~\cite{Doring:2011vk} and is
expected to be more efficient: the simultaneous fit of pure $P$-wave and $SP$-wave levels leads
to smaller uncertainties in the infinite-volume limit than the two-step procedure discussed
before. 

 A generalization to the multiple partial wave, multiple channel situation, using the
corresponding matrices in channel- and partial wave-space as those of Eqs.~(\ref{vvvv},
\ref{gmat}), is in principle straightforward but will require very high precision lattice data;
see Ref.~\cite{Doring:2011vk} where this issue is discussed for the two-channel case.

As performed in Sec.~\ref{sec:results} for the $\sigma(600)$, there is no particular 
problem to generate pseudo-data for the $\kappa/K^*$-system and disentangle phase shifts
following the strategy formulated here. However, to not overload this study, we concentrate on
the somewhat simpler case of the $\sigma(600)$ in the next section.


\subsection{Results}
\label{sec:results}
\begin{figure*}
\begin{center}
\includegraphics[width=0.2117\textwidth]{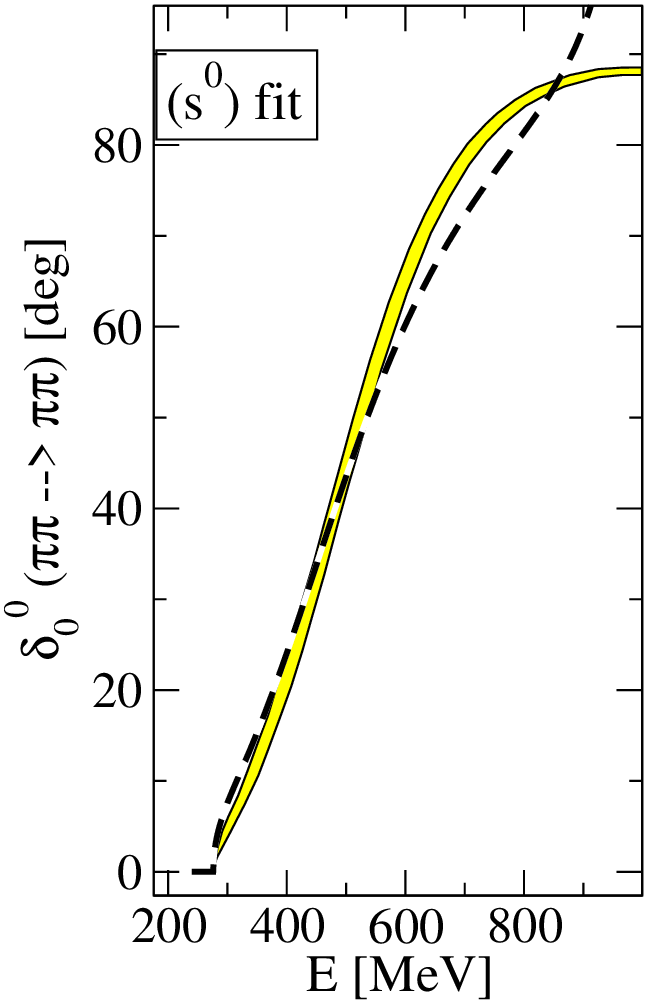}
\hspace*{-0.395cm}
\includegraphics[width=0.17\textwidth]{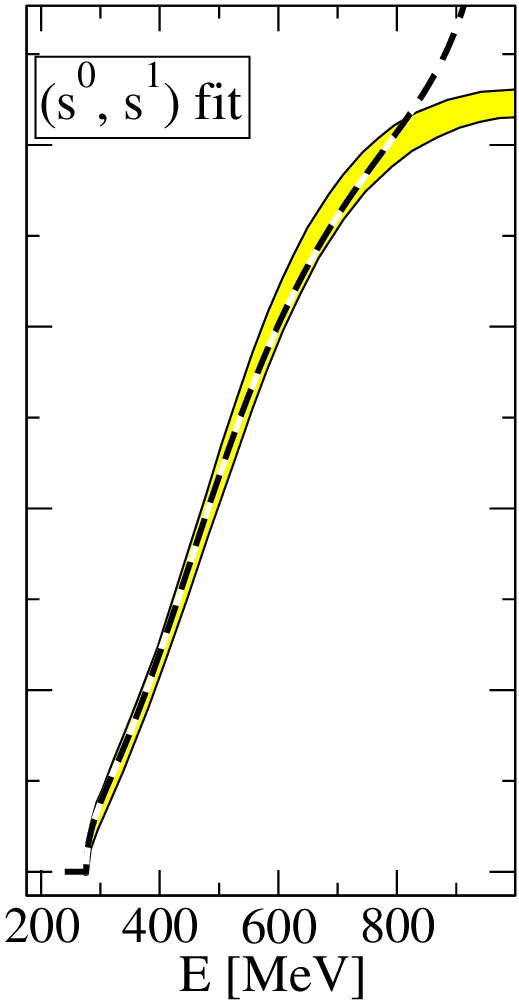} 
\hspace*{-0.395cm}
\includegraphics[width=0.17\textwidth]{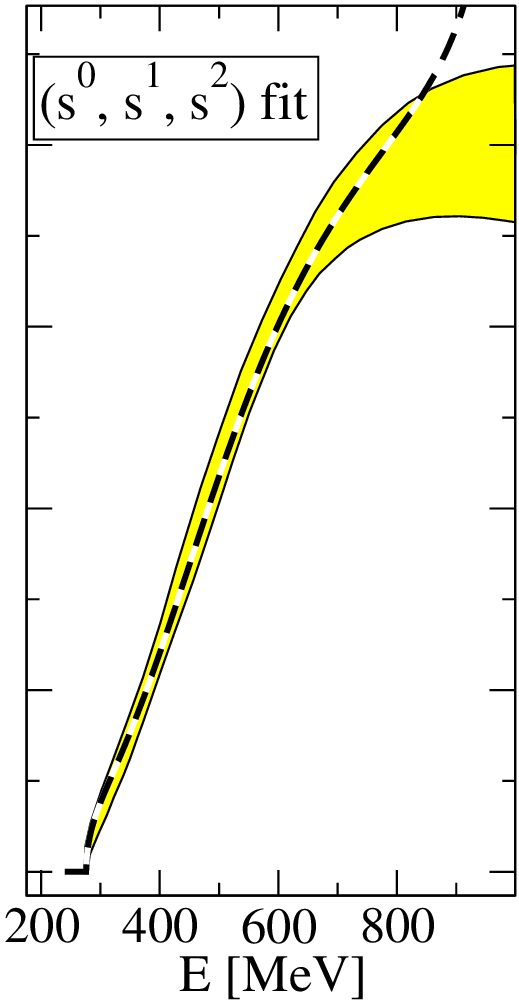}
\hspace*{-0.395cm}
\includegraphics[width=0.17\textwidth]{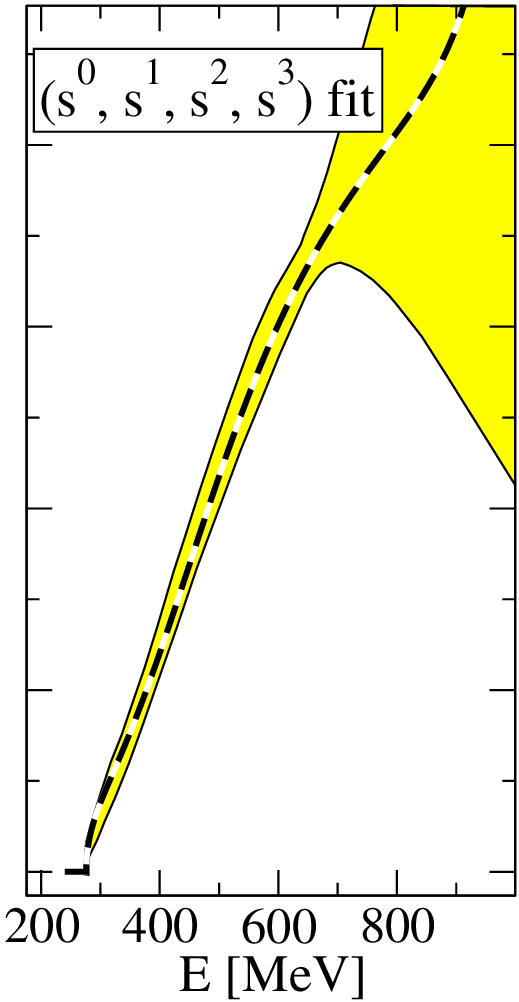} 
\end{center}
\caption{Extracted phase shifts (bands) using synthetic data points as indicated in
Fig.~\ref{fig:levels_sigma}, with a 10~MeV error for each data point. The actual phase shift is
shown with the dashed lines.
}
\label{fig:phases_sigma}
\end{figure*}

\begin{figure}
\begin{center}
\includegraphics[width=0.47\textwidth]{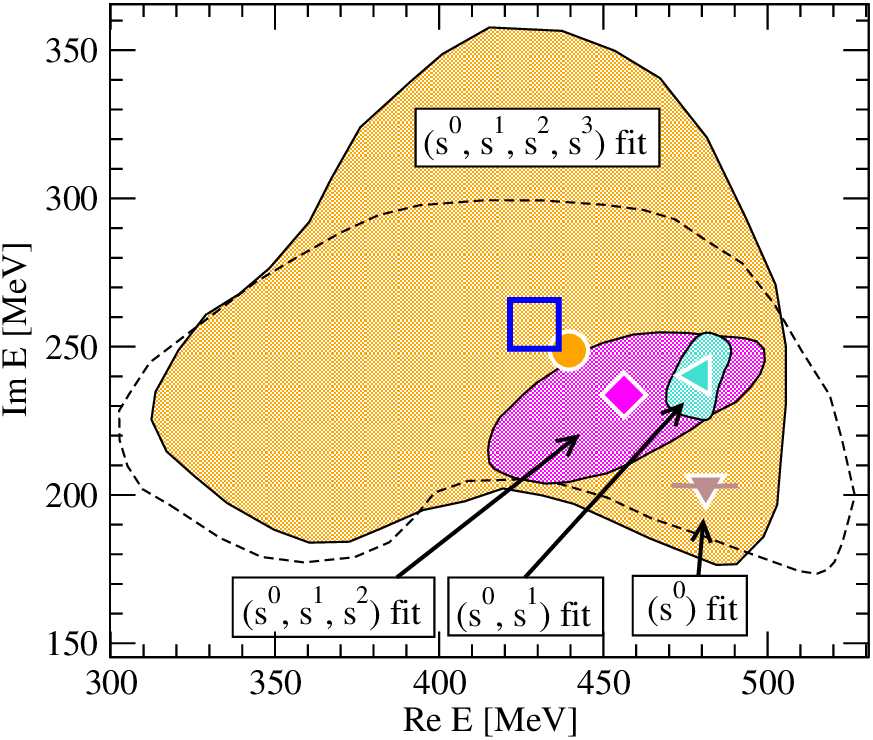} 
\end{center}
\caption{Pole of the $\sigma(600)$ extracted from the synthetic data shown in
Fig.~\ref{fig:levels_sigma}.  The actual pole position is indicated with the large square.
Extracted pole positions for the different fits: triangle down: $(s^0)$ fit; triangle left:
$(s^0,s^1)$ fit; diamond:  $(s^0,s^1,s^2)$ fit; circle: $(s^0,s^1,s^2,s^3)$ fit. Also,
uncertainties are shown (shaded areas). The dashed line shows the uncertainty from
Ref.~\cite{Doring:2011nd}, where -- instead of two volumes and five boosts as done here -- 6
different volumes had to be considered.}
\label{fig:poles_sigma}
\end{figure}

The analysis of the $\sigma(600)$ proceeds as described following Eq.~(\ref{vfit}) and in
Ref.~\cite{Doring:2011nd}, with the fit potential $V^{\rm fit}$ from Eq.~(\ref{vfit}). The fits
are labeled according to the powers of $s$ used in $V^{\rm fit}$. Once a minimum is found,
parameter errors are determined. The parameter error for a parameter $a$ is defined by the range
of $a$ in which $\chi^2<\chi^2_{\rm best}+1$, under the constraint that all other parameters are
optimized. Then, within the errors, random parameter sets are generated and only those sets kept
for which $\chi^2<\chi^2_{\rm best}+1$. For each of these sets, phase shift and pole position are
calculated. The resulting bands and areas, for phase and pole position, respectively, are shown
in Figs.~\ref{fig:phases_sigma} and \ref{fig:poles_sigma}. Note that the uncertainty area  for
the pole position in the $(s^0)$ fit, shown in Fig.~\ref{fig:poles_sigma}, shrinks to a line. For
the best $\chi^2$, pole positions are indicated with symbols in Fig.~\ref{fig:poles_sigma}. The
actual phase shift and pole position, derived from the hadronic interaction that was used to
generate the synthetic data, are also indicated in the figures.

As visible for phase shifts and pole positions, fits with larger number of free parameters
result (trivially) in larger uncertainties, and, of course, in a better $\chi^2$. For the
central values of the pole positions, we observe that with an increasing number of parameters,
the actual and the fitted pole positions get closer, but even the 4-parameter fit does not
perfectly match the actual pole position although it has the best $\chi^2$ of all fits. As
discussed in Sec.~\ref{fig:poles_sigma}, there are terms of higher order in $s^N$, $N>3$, in the
original potential that cause this small but finite discrepancy. In the fit of actual lattice
data, one will have the same effect, of course. 

As discussed in Ref.~\cite{Doring:2011nd} the effect of the heavier channel -- given by $\bar
KK$ in the present case -- usually can be well absorbed in the coefficients of the expansion of
the fit potential, but as the $\sigma(600)$ is very broad, small remaining discrepancies become
large far in the complex plane. The same behavior was found in Ref.~\cite{Doring:2011nd}, where
$L$ was varied to extract the $\sigma(600)$ pole. The situation cannot be improved by
explicitly including the $\bar KK$ channel in the extraction process, in contrast to the case
of the $f_0(980)$ where this is possible and necessary~\cite{Doring:2011vk}. Here, i.e. still
far below the $\bar KK$ threshold, the channel transitions $V_{\pi\pi\to\bar K K}$ and $V_{\bar
K K\to\bar K K}$ are very weakly constrained. The large number of new free parameters, tied to
these additional transitions, would immediately lead to drastically increased uncertainties on
the observables and large parameter correlations as has been tested.

We observe, in any case, that using synthetic data from values of $L$ smaller than
$2.5M_\pi^{-1}$, i.e. higher energies $E$, immediately helps to narrow down the uncertainties
of the phase shifts at higher energies, shown in Fig.~\ref{fig:phases_sigma}. However, in that
case the central value of the pole position in the $(s^0,s^1,s^2,s^3)$ fit, shown with the
circle in Fig.~\ref{fig:poles_sigma}, starts to deviate considerably from the actual pole
position (empty square), which is a sign of the increasing effect of the $\bar KK$ channel on
the lattice data closest to the $\bar KK$ threshold. Choosing box sizes of $L=2.5M_\pi^{-1}$
and $L=3.5M_\pi^{-1}$ as done here provides, thus, a good compromise. These values of $L$, for
which partial wave mixing can be safely neglected, is promising for the setup of an actual
lattice simulation to extract the $\sigma(600)$.

Finally, we would like to compare the present results to those of Ref.~\cite{Doring:2011nd}
where -- instead of two required volumes as is the case here -- 6 different volumes had to be
considered, implying a much increased numerical effort for actual lattice calculations. The
uncertainty of the $\sigma(600)$ pole position, coming from synthetic data with the same 10~MeV
error as used here, is shown with the dashed line in Fig.~\ref{fig:poles_sigma}. The extension
is of similar size as the one of the $(s^0,s^1,s^2,s^3)$ fit performed here.

Thus, with only two different volumes and using data from five boosts for each, one can expect
results of similar precision than from six different values of $L$ without boost. This
demonstrates that the proposed extraction method can be quite effective in the analysis of
actual lattice data.


\subsection{Statistical error}

\begin{figure}
\begin{center}
\includegraphics[width=0.47\textwidth]{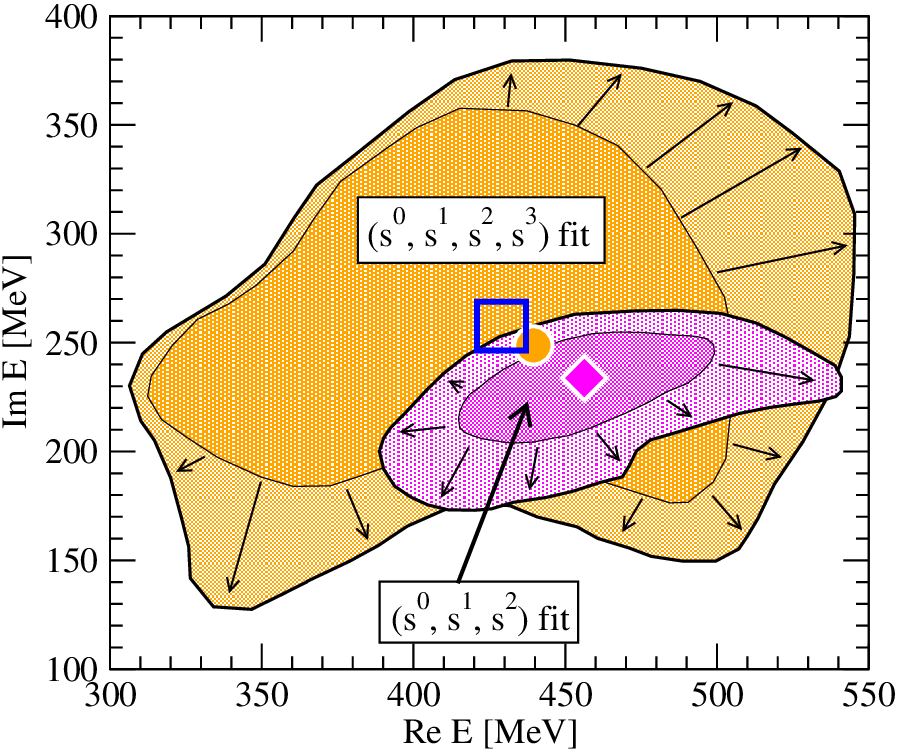} 
\end{center}
\caption{Uncertainties of the $\sigma(600)$ pole position from the statistical uncertainty of
the data. As indicated with the arrows, the uncertainties increase compared to those of
Fig.~\ref{fig:poles_sigma}. The cases of the $(s^0,\,s^1,\,s^2)$ and the
$(s^0,\,s^1,\,s^2,\,s^3)$ fits are shown.}
\label{fig:poles_sigma_incr}
\end{figure}

One should note that for the generation of the synthetic data we have so far only assigned an
error, but not allowed the statistical fluctuation of the centroids of the error bars.  In other
words, we have assumed that the entire error is systematic. Here,  we allow, in addition to the
10~MeV error, a statistical fluctuation of the centroids by 5~MeV, as has been done in
Ref.~\cite{Doring:2011vk} for the extraction of the $f_0(980)$. The values of 10~MeV for the
error bar and 5~MeV for its fluctuation are chosen to demonstrate the  effect; when it comes to
the analysis of actual lattice data, these values and in particular  the fraction of the
statistical in the total error have to be adapted, of course. Using the same procedure as in
Ref.~\cite{Doring:2011vk}, we can estimate the resulting uncertainties, shown in
Fig.~\ref{fig:poles_sigma_incr} for two fits.
As the figure shows, the uncertainties increase by around one third for the chosen values. 


\section{Summary}
The present study provides a formulation for the scattering of two particles confined in a 
finite box with total nonzero momentum, adapted to the chiral unitary framework.  The idea is
based on extending previously known techniques for zero momentum, discretizing the energy
levels by imposing the boundary conditions in the moving frame. 

Given a hadronic interaction, levels for the first five boosts $\vec P=(2\pi/L)(0,0,0)$ to
$(0,0,2)$  can be predicted and attributed to the subgroups of cubic symmetry. Employing
coupled-channel  unitarized chiral  perturbation theory including NLO terms, we derive the levels
for the mixed-partial wave system with $I=1/2$, $S=-1$ and $L=0,1$ [$\kappa(800)$ and $K^*(892)$,
respectively] as well as for the scalar sector with $I=0$, $S=0$ and $L=0,2$ where the
$\sigma(600)$ resides. 

We demonstrate for the $\kappa(800)/K^*(892)$ system that partial wave  mixing is a very large
effect for realistic box sizes and needs to be taken into account.  To disentangle the $S$-wave
from $P$- or $D$-wave, we derive a set of  equations in the one-channel formalism that are
shown to be very precise as long as one stays below the inelastic thresholds. 

Furthermore, we present a scheme in which the hadronic interaction is expanded in energy to
allow  for the extraction of the infinite volume limit, simultaneously fitting levels for
different boosts  and at different energies.  The model-independent information from the lowest
order in the chiral expansion is kept explicitly in this expansion, greatly stabilizing the fit
to lattice data.  Such statistical analyses can be used for actual lattice data,  or, as done
here, serve to determine promising lattice setups and the  accuracy of lattice data to allow
for reliable resonance extraction.

The method is tested for the example of the $\sigma(600)$. First,  we show that for
$L>2.5\,M_\pi^{-1}$ effects from $SD$-wave mixing can be neglected. Second, synthetic
lattice-data are produced and analyzed. We find that with only two different box sizes one can
expect a similar  precision on the $\sigma(600)$ pole position as by varying as much as 6
different box sizes at zero total momentum. 

Using information from moving frames is, thus, indeed rewarding since, with only a few
different box sizes, phase shifts and resonance parameters of excited mesons can be determined.
 

\section*{Acknowledgments}
 We would like to thank Sasa Prelovsek and Luis Roca for useful discussions. This work is partly
supported by DGICYT contracts  FIS2006-03438, the Generalitat Valenciana in the program Prometeo
and  the EU Integrated Infrastructure Initiative Hadron Physics3 Project under Grant Agreement
no. 283286. We also acknowledge the support by DFG (SFB/TR 16, ``Subnuclear Structure of
Matter''), by COSY FFE under contract 41821485 (COSY 106) and the Sino-German CRC 110 ``Symmetries and
the Emergence of Structure in QCD. A.R. acknowledges  support of the
Shota Rustaveli National Science Foundation  (Project DI/13/02-100/11).	


\appendix
\section{Parameterization of the $D$-wave from Ref.~\cite{GarciaMartin:2011cn}}
\label{appe}
For convenience we quote the parameterization of the isospin zero $D$-wave from
Ref.~\cite{GarciaMartin:2011cn}. To construct the $\pi\pi$ $D$-wave to $D$-wave transition
$V_D^{(2,2)}$, we express it in terms of $\cot\delta_D$ given in Eq.~(A9) of
Ref.~\cite{GarciaMartin:2011cn},
\be
[V_D^{(2,2)}]^{-1}={\rm Re}\,G-\frac{p\,\cot\delta_D}{8\pi E} \ ,
\ee
valid up to the $\bar KK$ threshold ($E< 2M_K$). Here, 
\ba
\cot\delta_D&=&
  \frac{E}{2p^5}\,\left(M_{f_2}^2-s\right)\,M_\pi^2\left[B_0+B_1\,w(s)\right] \ , \non
w(s)&=&\frac{E-\sqrt{E_0^2-E^2}}{E+\sqrt{E_0^2-E^2}},\, E_0=1050\,\text{MeV}
\ea
where $M_{f_2}=1275.4$~MeV, $B_0=12.40$, $B_1=10.06$, and $p$ is the relative momentum of the
pions.


\end{document}